\documentclass[reprint,superscriptaddress,amsmath,amssymb,aps,prb]{revtex4-2}
\usepackage{xcolor}
\usepackage{graphicx}
\usepackage{dcolumn}
\usepackage{bm}
\usepackage{hyperref}
\usepackage[normalem]{ulem}
\usepackage[mathlines]{lineno}

\begin{document}

\title{Dipolar and quadrupolar spin supersolid states in a spin-1 triangular antiferromagnet}

\author{Yixuan Huang}
\affiliation{RIKEN Center for Emergent Matter Science (CEMS), Wako 351-0198, Japan}

\author{Yuan Gao}
\affiliation{Institute of Theoretical Physics, Chinese Academy of Sciences, Beijing 100190, China}

\author{Wei Li}
\affiliation{Institute of Theoretical Physics, Chinese Academy of Sciences, Beijing 100190, China}

\author{Seiji Yunoki}
\affiliation{RIKEN Center for Emergent Matter Science (CEMS), Wako 351-0198, Japan}
\affiliation{RIKEN Center for Computational Science (R-CCS), Kobe 650-0047, Japan}
\affiliation{RIKEN Center for Quantum Computing (RQC), Wako 351-0198, Japan}
\affiliation{RIKEN Pioneering Research Institute (PRI), Wako 351-0198, Japan}

\author{Sadamichi Maekawa}
\affiliation{RIKEN Center for Emergent Matter Science (CEMS), Wako 351-0198, Japan}
\affiliation{Advanced Science Research Center, Japan Atomic Energy Agency, Tokai 319-1195, Japan}

\date{\today}

\begin{abstract}
We present a systematic numerical study of the spin-1 antiferromagnetic Heisenberg model on the triangular lattice in an out-of-plane magnetic field, using Density Matrix Renormalization Group (DMRG) methods. By mapping out the quantum phase diagram as a function of the single-ion anisotropy $D_z$ and magnetic field, we identify distinct dipolar and quadrupolar spin supersolid states, characterized by spontaneous U(1) symmetry breaking with finite spin superfluid stiffness coexisting with longitudinal translational symmetry breaking. At zero field, the dipolar spin supersolid with a `Y'-type spin configuration persists down to $D_z = 0$, whereas the quadrupolar spin supersolid prevails at large $D_z$. At intermediate fields, the phase diagram is dominated by an up-up-down phase. At high fields below saturation, a quadrupolar spin superfluid emerges in the large-$D_z$ regime, whereas a dipolar spin supersolid with a `V'-type spin configuration dominates at small $D_z$. These phases are characterized through their order parameters and spin superfluid stiffness using calculations on various system sizes. Furthermore, the dynamical spin structure factor is obtained across the phase diagram, where characteristic spectral signatures of different phases are observed, including the gapless Goldstone mode and the roton-like minima. These features are directly accessible to inelastic neutron scattering experiments. Our results provide a theoretical understanding of the interplay between frustrations, anisotropy, and Zeeman interactions in driving distinct spin supersolid phases in the spin-1 system, which are relevant to various triangular-lattice antiferromagnets such as Na$_2$BaNi(PO$_4$)$_2$ and K$_2$Ni(SeO$_3$)$_2$.
\end{abstract}
\maketitle

\section{Introduction}
The supersolid is an exotic quantum phase of matter that simultaneously exhibits off-diagonal long-range order associated with superfluidity and diagonal long-range order associated with translational symmetry breaking. This intriguing concept was originally proposed in the context of solid helium~\cite{leggett1970can, chester1970speculations}. Early experiments reported signatures of supersolidity in $^4$He~\cite{kim2004probable}, triggering intense theoretical and experimental interest. However, subsequent investigations suggested that the observed signals were most likely attributable to disorder effects or dislocations rather than intrinsic supersolidity~\cite{balibar2010enigma, chan2013overview}. To date, no definitive experimental evidence for a supersolid phase in solid helium has been established.

In parallel, the concept of supersolidity has been extended to other strongly correlated systems~\cite{boninsegni2012colloquium}. Ultracold atoms in optical lattices provide clean and controllable platforms where lattice supersolids have been observed~\cite{bloch2008many,landig2016quantum,sinha2025supersolid}, and more recently, ultracold quantum gases with dipolar interactions have yielded direct experimental evidence for supersolidity~\cite{tanzi2019observation, bottcher2019transient, chomaz2019long, guo2019low, tanzi2019supersolid, natale2019excitation,norcia2021two, tanzi2021evidence, norcia2022can, recati2023supersolidity, vsindik2024sound, bougas2026signatures}. Beyond atomic systems, frustrated quantum magnets offer an alternative, material-based route to realizing supersolid physics. In these systems, the mapping between bosons and quantum spins allows supersolidity to manifest as a spin supersolid (SS)~\cite{jiang2009supersolid}, whose characteristic signatures can, in principle, be probed by inelastic neutron scattering (INS) and thermodynamic measurements in realistic magnetic compounds.

Among frustrated quantum magnets, triangular-lattice antiferromagnets have emerged as particularly promising platforms for the study of spin supersolidity~\cite{huang2026emergent}, which includes several compounds with effective spin-$\frac{1}{2}$~\cite{ono2026microscopic}, such as $\text{Na}_{2}\text{BaCo}(\text{PO}_{4})_{2}$~\cite{sheng2022two,xiang2024giant,gao2024double,mou2024comparative,zhang2025field,hussain2025experimental,popescu2025zeeman,xu2025nmr,woodland2025continuum,sheng2025continuum,ferreira2026direct}, $\text{K}_{2}\text{Co} (\text{SeO}_{3})_{2}$~\cite{zhong2020frustrated,zhu2024continuum,Zhu2025wannier,chen2026phase,zhang2026nonperturbative}, and $\text{Rb}_{2}\text{Co} (\text{SeO}_{3})_{2}$~\cite{zhong2020frustrated,shi2025absence,cui2026spin}. Geometric frustration and competing interactions on the triangular lattice give rise to a rich variety of exotic phases. In the presence of an applied magnetic field, the interplay between frustration, quantum fluctuations, and Zeeman coupling can drive a series of quantum phase transitions, giving rise to magnetization plateaus and, crucially, to SS states in which spin superfluidity coexists with translational symmetry-breaking magnetic order. These triangular antiferromagnets therefore provide an ideal setting for bridging theory and experiment.

Extensive numerical studies have been carried out on the spin-$\frac{1}{2}$ triangular-lattice Heisenberg antiferromagnet, mapping out its quantum phase diagrams and establishing the SS phases~\cite{heidarian2010supersolidity,chen2013ground,yamamoto2014quantum,starykh2015unusual, sellmann2015phase,chi2024dynamical,gao2022spin,gao2024double,gao2025spin,xu2025simulating,sheng2025continuum,ulaga2025easy,gallegos2025phase,flores2025unconventional,keselman2025j1,bose2025modified,ulaga2026anisotropic,huang2026dissipationless,kadosawa2026nontrivial,mauri2026nonlinear}. However, the spin-1 Heisenberg model remains much less explored, with previous studies limited to cluster mean field theory~\cite{moreno2014case}, perturbative approaches~\cite{seifert2022phase}, and models relevant to Na$_2$BaNi(PO$_4$)$_2$~\cite{sheng2025bose,sheng2025possible,huang2025universal}. The spin-1 model is qualitatively different from its spin-$\frac{1}{2}$ counterpart because it supports higher-order spin structures such as the quadrupolar order, which may be promoted by single-ion anisotropy that is relevant in realistic materials. Indeed, Na$_2$BaNi(PO$_4$)$_2$ has been recently shown to realize strong single-ion anisotropy, with $D_z/J \approx 4$, which stabilizes a quadrupolar SS~\cite{sheng2025bose}, where out-of-plane translational symmetry breaking order coexists with an in-plane quadrupolar moment. Such quadrupolar order represents a form of ``hidden'' spin-nematic order, where the spin symmetry is spontaneously broken without any conventional dipolar order. Furthermore, INS experiments have provided evidence of magnon-pair condensation as the origin of the quadrupolar moment~\cite{sheng2025possible,huang2025universal}. However, the nature of the narrow low-energy dynamical excitations associated with this quadrupolar supersolid remains not fully understood. In addition, the spin-1 triangular antiferromagnet K$_2$Ni(SeO$_3$)$_2$ exhibits much weaker single-ion anisotropy~\cite{li2023k2ni}, placing it in a distinct parameter regime of the quantum phase diagram where SS phases involving dipolar and quadrupolar orders may emerge. Recent zero-field INS experiments have revealed gapless excitations near the $\mathbf{K}$ points and a broad excitation continuum at high energies, which has been suggested to originate from multiparticle excitations~\cite{kim2026spin}. Understanding how these dynamical excitations evolve under a magnetic field and across different phases represents an important open question.

\begin{figure}
\centering
\includegraphics[width=1\linewidth]{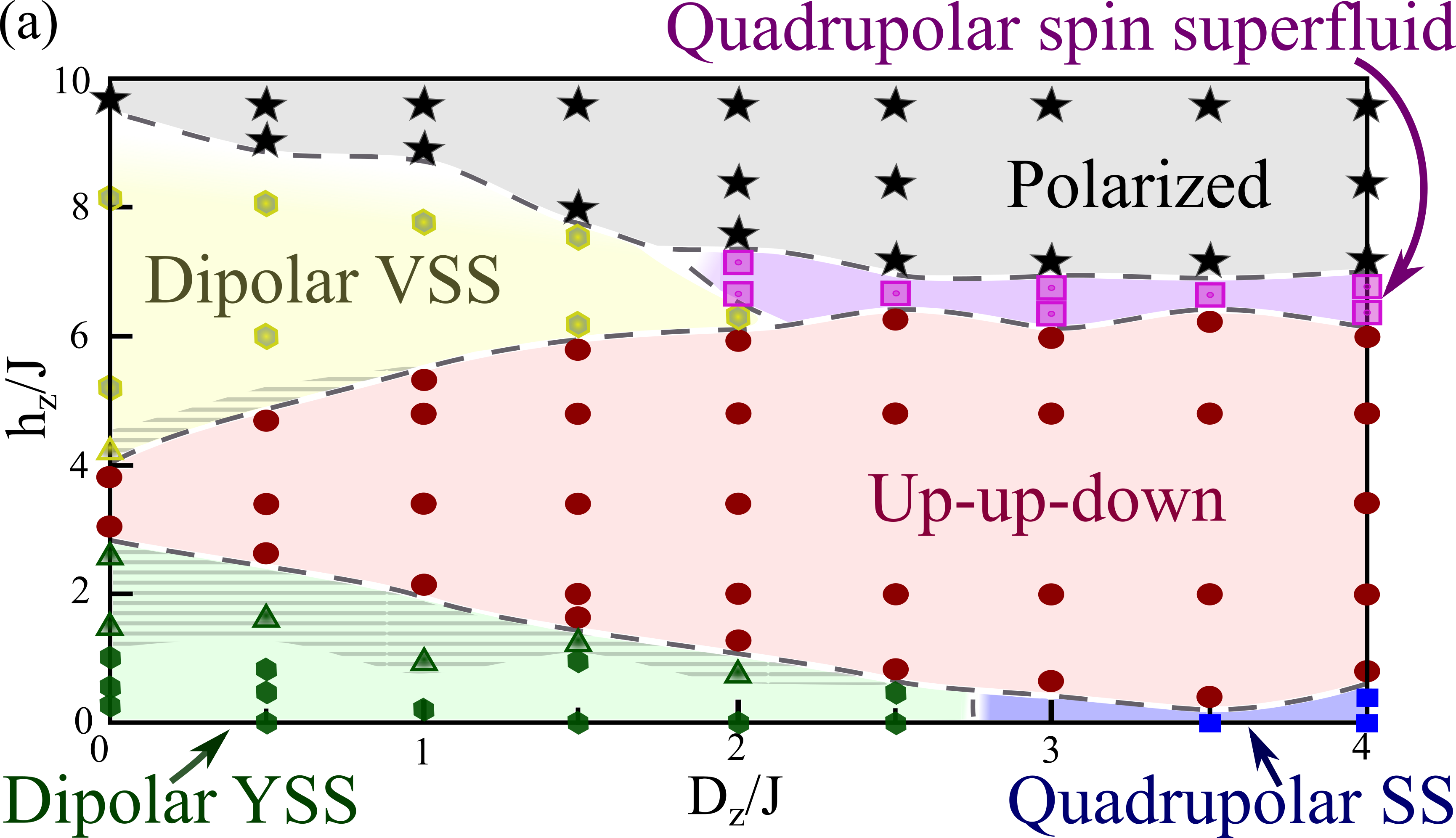}
\includegraphics[width=1\linewidth]{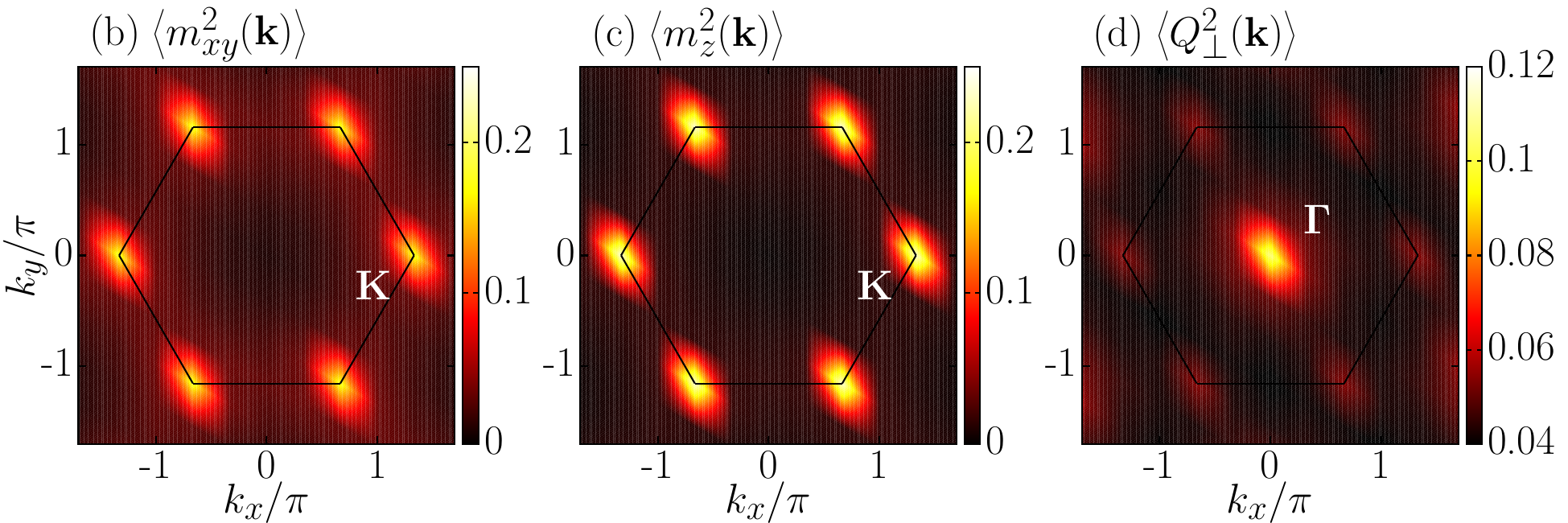}
\includegraphics[width=1\linewidth]{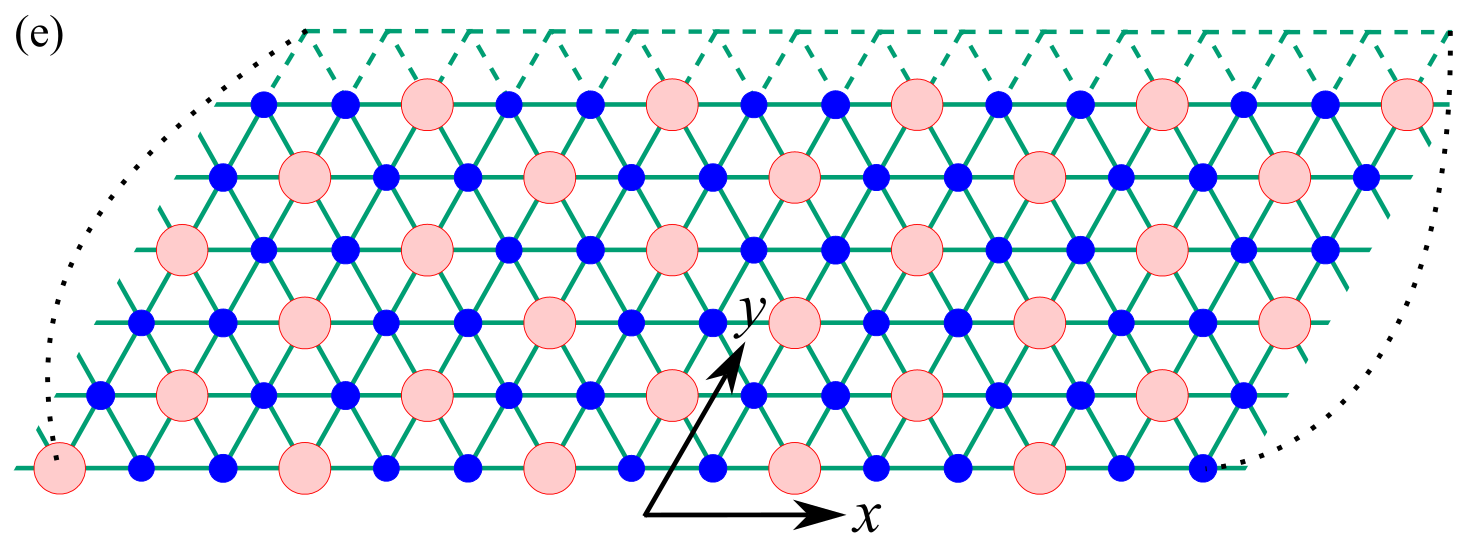}
\caption{Panel (a) sketches the quantum phase diagram with various single-ion anisotropy $D_z/J$ and magnetic field $h_{z}/J$. The symbols represent parameter points studied using DMRG methods. The shaded regimes indicate the vanish of quadrupolar order in the thermodynamic limit within the dipolar SS phases. Panels (b) and (c) show the transverse and longitudinal dipolar orders defined in Eq.~\ref{eq_dipolar_order}, respectively. Panel (d) shows the quadrupolar order defined in Eq.~\ref{eq_quadrupole_order}. Panel (e) shows the real-space distribution of $\left\langle S_{i}^{z}\right\rangle $. The blue solid circles represent positive $\left\langle S_{i}^{z}\right\rangle $, and red shaded circles represent negative $\left\langle S_{i}^{z}\right\rangle $ with radius proportional to its magnitude. The red shaded circles have $\left\langle S_{i}^{z}\right\rangle \approx -0.79$. In addition, the periodic boundary condition in the $y$ direction is illustrated in panel (e). Panels (b), (c), (d), and (e) are obtained at $D_{z}/J=2, h_{z}/J=0$ on the $N=36 \times 6$ lattice, and only the center part of the lattice is shown in panel (e).}
\label{Fig1_phase_diagram}
\end{figure}

Motivated by the recent discovery of the spin-1 triangular antiferromagnets Na$_2$BaNi(PO$_4$)$_2$~\cite{li2021quantum} and K$_2$Ni(SeO$_3$)$_2$~\cite{li2023k2ni}, we numerically study the spin-1 triangular-lattice antiferromagnetic Heisenberg model by tuning single-ion anisotropy $D_z$ and out-of-plane magnetic field $h_{z}$, using density matrix renormalization group (DMRG) methods~\cite{white1992density,white1993density,schollwock2011density}. A global phase diagram is mapped out as a function of $D_z$ and $h_{z}$ with distinct SS phases characterized by their order parameters and superfluid stiffness. At weak magnetic fields, we identify a dipolar SS—characterized by a dominant dipolar order $\left\langle m_{xy}^{2}(\mathbf{k}) \right\rangle$ and weak but finite quadrupolar order $\left\langle Q_{\perp }^{2} ( \mathbf{k }) \right\rangle$ in the thermodynamic limit—at small $D_{z}$, while the large-$D_z$ regime is dominated by a quadrupolar SS with finite $\left\langle Q_{\perp }^{2} ( \mathbf{k }) \right\rangle$ but vanishing $\left\langle m_{xy}^{2}(\mathbf{k}) \right\rangle$. With increasing magnetic field at small $D_z$, the quadrupolar order is progressively suppressed and eventually vanishes before the system enters the up-up-down (UUD) phase. At high magnetic fields, the dipolar SS dominates at small $D_z$, while a quadrupolar spin superfluid emerges in the large-$D_z$ regime which is characterized by finite quadrupolar order without out-of-plane translational symmetry breaking. Besides the order parameters, we calculate the spin superfluid stiffness which is directly associated with spin superfluid density and supercurrents~\cite{sonin2010spin}. Furthermore, we calculate the dynamical spin structure factor across the phase diagram, revealing sharp spectral signatures of the distinct SS phases. 

The quantum phases near $D_{z}=4$ are particularly relevant to Na$_2$BaNi(PO$_4$)$_2$, where the numerical results of the longitudinal dynamical spin structure factor at $D_{z}=4$ can be directly compared with INS experiments, providing insights into the nature of the narrow low-energy excitations observed in Na$_2$BaNi(PO$_4$)$_2$~\cite{sheng2025possible,huang2025universal}. More broadly, we map out the global phase diagram and study the quantum phase transitions between dipolar SS, quadrupolar SS, quadrupolar spin superfluid, UUD, and polarized phases. The results near $D_{z}=0$ may also be relevant to K$_2$Ni(SeO$_3$)$_2$ under an applied magnetic field where the dynamical spin structure factor can be compared with INS experiments~\cite{kim2026spin}.


The rest of this paper is organized as follows. Section~\ref{sec:model} introduces the spin-1 anisotropic Heisenberg model and describes the numerical methods. Section~\ref{sec:phase_diagram} presents the quantum phase diagram and characterizes various SS states through the order parameters and superfluid stiffness, which are distinct from the conventional UUD state. The phase transitions in zero field are also discussed. In Sec.~\ref{sec:SS_Dz0}, we focus on the $D_{z}=0$ limit under finite magnetic fields, identify different SS states [Sec.~\ref{subsec:SS_Dz0:SS}], their spin superfluid stiffness [Sec.~\ref{subsec:SS_Dz0:stiff}], and discuss the dynamical spin structure factors for each phase [Sec.~\ref{subsec:SS_Dz0:Dynamical}]. We then turn to the field-induced phases at $D_{z}=4$ in Sec.~\ref{sec:SS_Dz4}, which are characterized by their SS orders [Sec.~\ref{subsec:SS_Dz4:SS}], spin superfluid stiffness [Sec.~\ref{subsec:SS_Dz4:stiff}], and characteristic spectral features in the dynamical spin structure factor [Sec.~\ref{subsec:SS_Dz4:Dynamical}]. Finally, the results are summarized in Sec.~\ref{sec:summary}.

\section{Model and methods}
\label{sec:model}
We study the spin-1 antiferromagnetic Heisenberg model on a triangular lattice, where the Hamiltonian is defined as 
\begin{align}
\label{eq_H}
H &= J\sum\limits_{\left\langle ij\right\rangle
}(S^{x}_{i} S^{x}_{j}+S^{y}_{i} S^{y}_{j}+\Delta _{z}S^{z}_{i} S^{z}_{j}) \\
&- D_{z}\sum\limits_{i}(S^{z}_{i})^{2}- h_{z}\sum\limits_{i}S^{z}_{i}. \nonumber
\end{align}
Here $\left\langle ij\right\rangle$ refers to the nearest neighbor sites and $J$ is set to 1 as the energy unit. $D_{z}$ refers to the single-ion anisotropy and $h_{z}$ refers to the out-of-plane magnetic field. To directly compare with the results of $\text{Na}_{2}\text{BaNi}(\text{PO}_{4})_{2}$, we set $\Delta _{z}=1.06$ as determined in Ref.~\cite{huang2025universal} by fitting the magnon dispersions in the polarized and UUD state.

Ground states are obtained using finite U(1) DMRG methods~\cite{white1992density,white1993density,schollwock2011density}. As illustrated in Fig.~\ref{Fig1_phase_diagram} (e), we consider cylindrical geometries with open boundary conditions along the $x$ direction and periodic boundary conditions along the $y$ direction with $L_x$ and $L_y$ sites, respectively. The total number of sites is $N=L_{x}\times L_{y}$. 
We consider cylinders with circumferences up to $L_{y} = 12$ and keep bond dimensions of up to $M=5000$, with typical truncation error $\epsilon \lesssim 10^{-5}$. In addition, physical quantities are extrapolated to the infinite bond dimension limit to further reduce finite bond dimension errors.

Real-time evolution is implemented using the time-dependent variational principle (TDVP)~\cite{haegeman2011time,haegeman2016unifying,li2023tangent}. In practice, one-site TDVP methods are used with bond dimensions enlarged through global Krylov vectors~\cite{yang2020time}. Up to $M=1500$ is used to evolve the states to a maximum time of $70/J$ with a time step of $0.5/J$. Time evolution of the ground states is primarily performed on cylinders with $L_{y} = 6$.


\section{Quantum phase diagram}
\label{sec:phase_diagram}
Figure~\ref{Fig1_phase_diagram} (a) summarizes our main results in the quantum phase diagram of the spin-1 triangular-lattice antiferromagnetic Heisenberg model as a function of the single-ion anisotropy $D_z/J$ and magnetic field $h_z/J$. The phase diagram hosts a rich variety of quantum phases arising from the interplay between frustrated interactions and anisotropy in a magnetic field. To distinguish the SS states with `Y'- and `V'-type spin configurations, we denote them as YSS and VSS, respectively, in Fig.~\ref{Fig1_phase_diagram} (a). Representative real-space distribution of $\left \langle S^{z}_{i} \right \rangle$ in the dipolar YSS phase is shown in Fig.~\ref{Fig1_phase_diagram} (e), with additional results for other phases provided in Appendix~\ref{Apendix_real_space}. 

\begin{figure}
\centering
\includegraphics[width=1\linewidth]{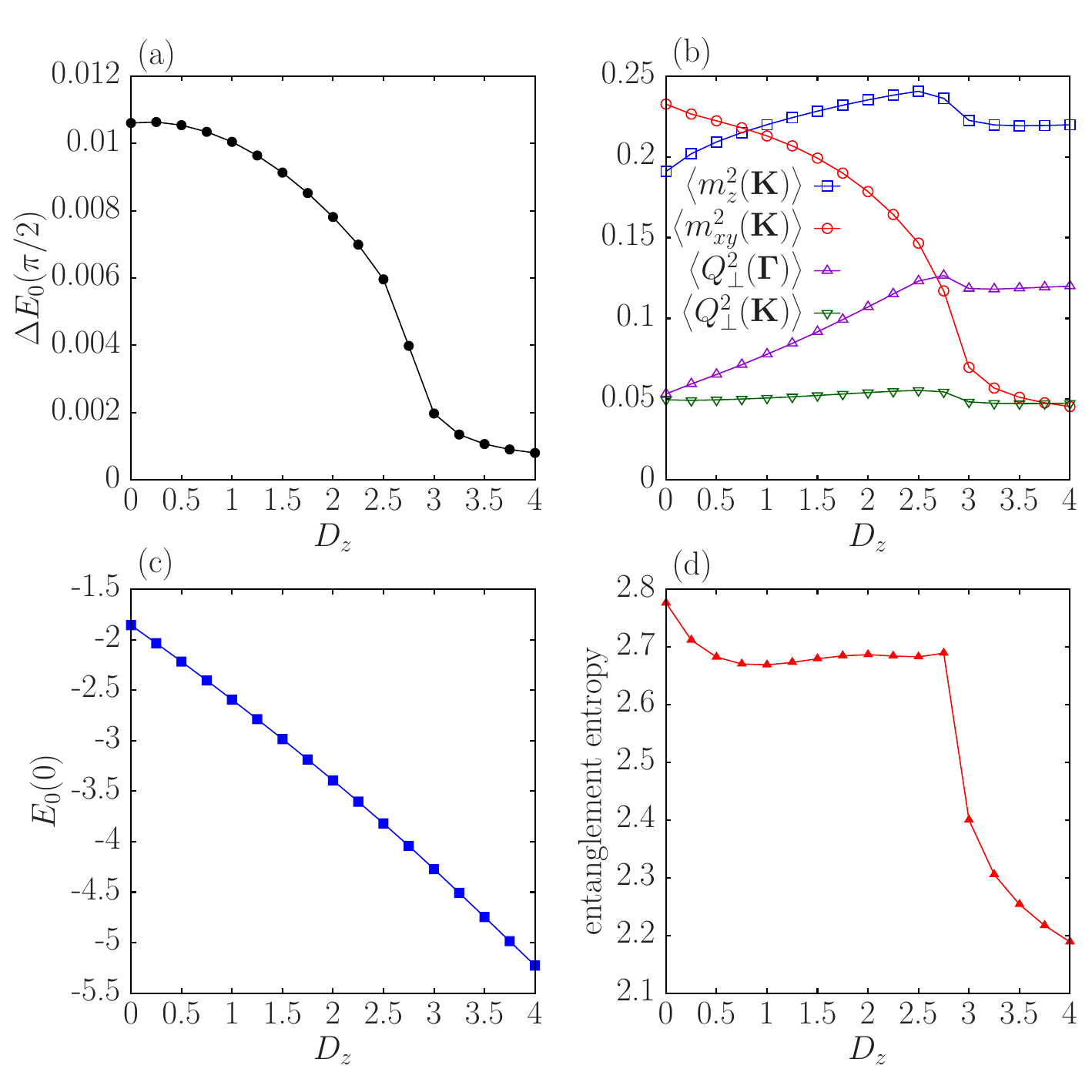}
\caption{(a) The superfluid stiffness, (b) various order parameters, (c) the ground-state energy per site, and (d) the entanglement entropy for various $D_{z}$ at $h_{z}= 0$. The results are obtained on a fixed lattice of $N = 36 \times 6$. We note that for larger $D_{z}> 1.75$ the ground state in the total $S^{z}=0$ sector converges to a stripe order with higher energy, which could be due to finite-size effect. In practice, the bulk properties of the ground state in the total $S^{z}=2$ sector are used to approximate the $h_{z}= 0$ limit where most magnetization is found to distribute near the open boundary in the $x$ direction.}
\label{Fig_zero_field_order}
\end{figure}

At zero $h_{z}$, the small-$D_{z}$ regime is occupied by the dipolar YSS, which is characterized by Bragg peaks in both the transverse spin structure factor $S^{xy}(\mathbf{k})$ and the longitudinal spin structure factor $S^{z}(\mathbf{k})$. The corresponding dipolar orders can be quantified by $\left\langle m_{xy}^{2} (\mathbf{k}) \right\rangle$ and $\left\langle m_{z}^{2} (\mathbf{k}) \right\rangle$ in the thermodynamic limit, which are defined as
\begin{equation}
\label{eq_dipolar_order}
\begin{split}
\left\langle m_{z}^{2}(\mathbf{k}) \right\rangle&=\frac{S^{z}(\mathbf{k})}{N^{\prime}}=\frac{1}{N^{\prime 2}}\sum_{i, j\in N^{\prime}} e^{i\mathbf{k}\cdot (\mathbf{r}_{i}-\mathbf{r}_{j})} \left\langle S^{z}_{i}S^{z}_{j}\right\rangle, \\
\left\langle m_{xy}^{2}(\mathbf{k}) \right\rangle&=\frac{S^{xy}(\mathbf{k})}{N^{\prime}} \\
&=\frac{1}{N^{\prime 2}}\sum_{i, j\in N^{\prime}} e^{i\mathbf{k}\cdot (\mathbf{r}_{i}-\mathbf{r}_{j})}\left\langle  S^{x}_{i}S^{x}_{j}+S^{y}_{i}S^{y}_{j}\right\rangle ,
\end{split}
\end{equation}
where $N^{\prime }=L_{y} \times L_{y}$ is chosen in the center of the lattice to minimize the open boundary effect. Representative results for $D_{z}/J=2$ are shown in Fig.~\ref{Fig1_phase_diagram} (b), where $\left\langle m_{xy}^{2}(\mathbf{k}) \right\rangle$ shows Bragg peaks at the $\mathbf{K}$ points. The real-space distribution of $\left \langle S^{z}_{i} \right \rangle$ shows a translational symmetry breaking pattern that is pinned by the open boundary [Fig.~\ref{Fig1_phase_diagram} (e)], with ``Y''-type spin configuration that corresponds to peaks in $\left\langle m_{z}^{2}(\mathbf{k}) \right\rangle$ at the $\mathbf{K}$ points [Fig.~\ref{Fig1_phase_diagram} (c)]. 

In addition, the transverse quadrupolar order $\left\langle Q_{\perp }^{2} (\mathbf{k}) \right\rangle$ shows prominent peaks at $\mathbf{\Gamma }$ points together with satellite peaks at $\mathbf{K}$ points, as shown in Fig.~\ref{Fig1_phase_diagram} (d). The quadrupolar order can be determined by $\left\langle Q_{\perp }^{2} (\mathbf{k}) \right\rangle$ in the thermodynamic limit, which is defined as
\begin{equation}
\label{eq_quadrupole_order}
\begin{split}
\left\langle Q_{\perp }^{2} (\mathbf{k}) \right\rangle =&\frac{Q^{\perp}(\mathbf{k})}{N^{\prime}} \\
=&\frac{1}{N^{\prime 2}}\sum_{i, j\in N^{\prime}} e^{i\mathbf{k}\cdot (\mathbf{r}_{i}-\mathbf{r}_{j})}\frac{1}{2} \left\langle  S^{+}_{i}S^{+}_{i}S^{-}_{j}S^{-}_{j}+\mathrm{H.c.}\right\rangle ,
\end{split}
\end{equation}
where $N^{\prime }$ is defined similarly. Because the numerical results are obtained on finite-size systems with finite bond dimensions that control the numerical truncation error, the results of $\left\langle m_{z }^{2} (\mathbf{k}) \right\rangle$, $\left\langle m_{xy}^{2} (\mathbf{k}) \right\rangle$, and $\left\langle Q_{\perp }^{2} ( \mathbf{k}) \right\rangle$ are first extrapolated to the infinite bond dimension limit for each system size, which is followed by the finite-size scaling to determine the order parameters in different phases. More details are provided in Appendix~\ref{Apendix_convergence}.

In addition, the superfluid stiffness $\rho_{s}$ can be probed by imposing a twisted $y$-boundary condition which is given by~\cite{jiang2009supersolid}
\begin{equation}
\label{eq_stiffness}
\rho_{s} \propto E_{0}(\theta)-E_{0}(0)\equiv \Delta E_{0}(\theta),
\end{equation}
where $E_{0}(\theta)$ denotes the ground-state energy per site with a twist angle $\theta$ that is applied to the spin flip terms across the periodic $y$ boundary according to $S_{i}^{+}S_{j}^{-}\rightarrow e^{i\theta }S_{i}^{+}S_{j}^{-}$; the illustration of the boundary bonds in the $y$ direction is shown in Fig.~\ref{Fig1_phase_diagram} (e).

\begin{figure*}
\centering
\includegraphics[width=0.95\linewidth]{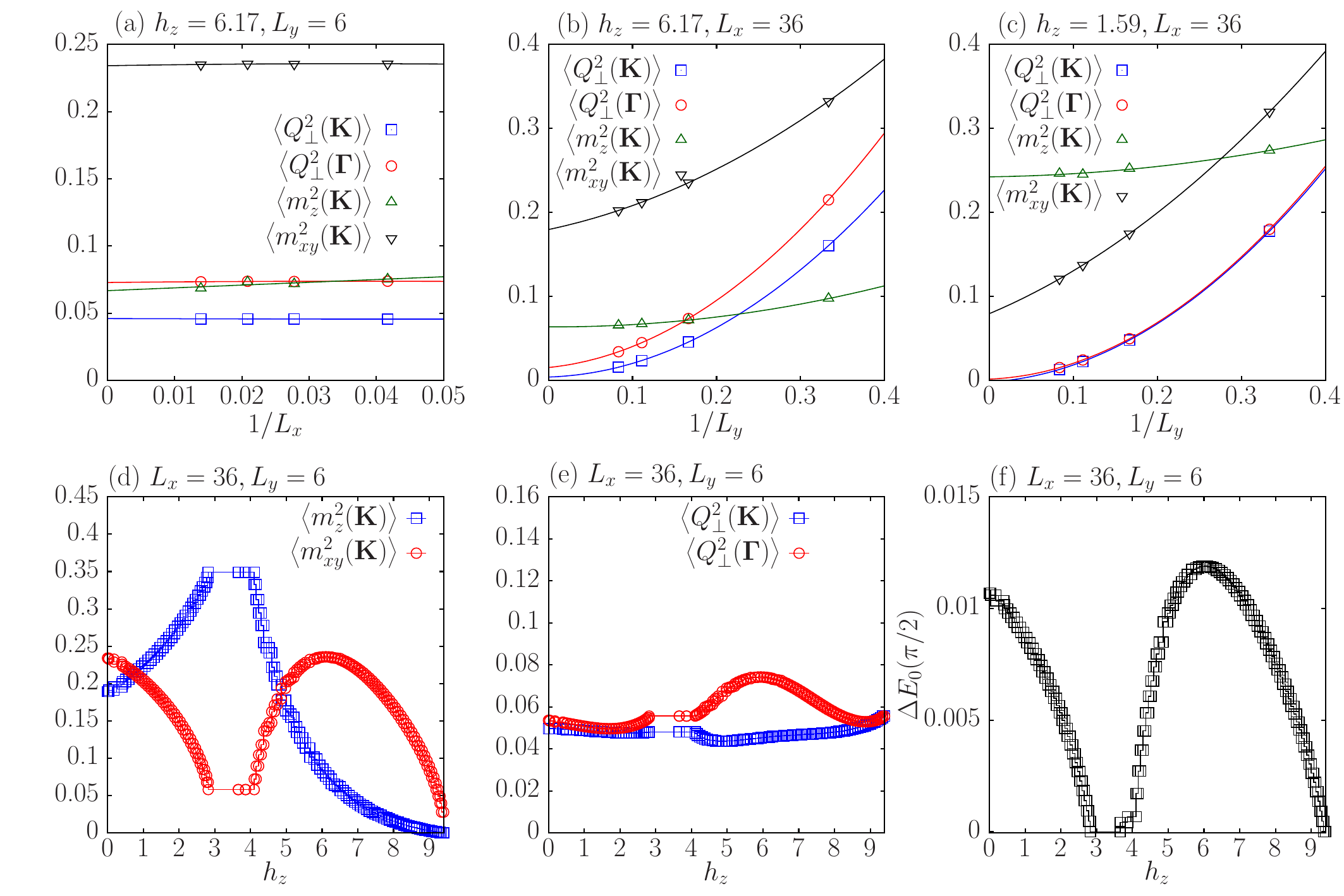}
\caption{Panel (a) and panels (b-c) show the finite-size scaling in $L_{x}$ and $L_{y}$ for various orders, respectively. Panels (d), (e), and (f) show the evolution of various orders and superfluid stiffness for different magnetic fields $h_{z}$ on a fixed lattice size. All results are obtained at $D_{z}=0$.}
\label{Fig_Dz0}
\end{figure*}

The quantum phase transition for various $D_z$ at $h_{z} = 0$ is shown in Fig.~\ref{Fig_zero_field_order}. At small $D_z$, the ground state is the dipolar YSS phase. The spin superfluid stiffness is probed using $\Delta E_{0}(\pi / 2)$ to capture contributions from both dipolar and quadrupolar orders, because quadrupolar order responds effectively to $2\theta $ while dipolar order responds to $\theta $. As shown in Fig.~\ref{Fig_zero_field_order} (a), $\Delta E_{0}(\pi / 2)$ decreases rapidly as $D_{z}$ increases to $2.75$, indicating a phase transition to the quadrupolar SS. This is consistent with the rapid suppression of $\left\langle m_{xy}^{2} (\mathbf{K}) \right\rangle$ at the same $D_{z}$, as shown in Fig.~\ref{Fig_zero_field_order} (b). In contrast, the longitudinal dipolar order $\left\langle m_{z}^{2} (\mathbf{K}) \right\rangle$ remains finite in the investigated $D_{z}$ range. The quadrupolar order $\left\langle Q_{\perp }^{2} ( \mathbf{\Gamma}) \right\rangle$ increases with increasing $D_{z}$ and tends to saturate for $D_{z} > 2.75$, while $\left\langle Q_{\perp }^{2} ( \mathbf{K}) \right\rangle$ remains small. In addition, the ground-state energy per site evolves smoothly across the phase transition, while the entanglement entropy shows a sharp decrease, as shown in Figs.~\ref{Fig_zero_field_order} (c) and (d), respectively. The smooth evolution of $\left\langle m_{xy}^{2} (\mathbf{K}) \right\rangle$, $\Delta E_{0}(\pi / 2)$, and the ground-state energy is consistent with a continuous phase transition.

The nature of the ground state at $h_{z} =0$ in the large $D_{z}$ limit could be understood through the perturbation expansion of $J/D_{z}$~\cite{sheng2025bose,sheng2025possible}. After projecting out the high energy state $|S^{z}=0>$, the resulting effective two-level model becomes
\begin{align}
\label{eq_effective}
H_{\textit{eff}} &= -\frac{J^{2}}{D_{z}}\sum\limits_{\left\langle ij\right\rangle
}(s^{x}_{i} s^{x}_{j}+s^{y}_{i} s^{y}_{j}+\tilde{\Delta }_{z}s^{z}_{i} s^{z}_{j}) \\
&- h_{z}\sum\limits_{i}s^{z}_{i}, \nonumber
\end{align}
where $\tilde{\Delta }_{z}=-2(1+2\frac{\Delta _{z} D_{z}}{J})$. The effective model shows a ferromagnetic coupling in the $xy$ component and an antiferromagnetic coupling in the $z$ component, giving rise to prominent peaks in $\left\langle Q_{\perp }^{2} ( \mathbf{k }) \right\rangle$ at $\mathbf{\Gamma }$ points and peaks in $\left\langle m_{z}^{2} (\mathbf{k}) \right\rangle$ at $\mathbf{K}$ points, respectively. However, the effective description is valid only for weak magnetic fields at large $D_{z}$, and we mainly rely on the DMRG results to establish the phase diagram at generic $h_{z}$ and $D_{z}$.

As shown in Fig.~\ref{Fig1_phase_diagram} (a), the dipolar YSS extends to finite magnetic fields. We note that within the shaded regime of dipolar YSS phase, $\left\langle Q_{\perp }^{2} ( \mathbf{\Gamma }) \right\rangle$ vanishes in the thermodynamic limit; further details are provided in Sec.~\ref{subsec:SS_Dz0:SS} and Appendix~\ref{Apendix_finite_size}. With increasing magnetic field, the system enters the UUD phase, which is characterized by a $1/3$ magnetization plateau and three-sublattice magnetic order. This phase is labeled as up-up-down in Fig.~\ref{Fig1_phase_diagram} (a). Upon further increasing the magnetic field, the system transits into the dipolar VSS phase. The dipolar VSS also exhibits weak but finite $\left\langle Q_{\perp }^{2} ( \mathbf{\Gamma }) \right\rangle$, except within the shaded regime. At small $D_{z}$ the dipolar YSS and VSS occupy an extended regime at low and high fields, respectively. However, at larger $D_{z}$, the dipolar YSS evolves into a quadrupolar SS with vanishing $\left\langle m_{xy}^{2} (\mathbf{K}) \right\rangle$, whereas the dipolar VSS evolves into a quadrupolar spin superfluid, in which both $\left\langle m_{xy}^{2} (\mathbf{K}) \right\rangle$ and $\left\langle m_{z}^{2} (\mathbf{K}) \right\rangle$ vanish, as the growth of single-ion anisotropy suppresses the $|S^z = 0\rangle$ occupancy and consequently the dipolar degrees of freedom. Phase transitions into the polarized state are found at sufficiently large $h_{z}$.


The quadrupolar order $\left\langle Q_{\perp }^{2} ( \mathbf{k}) \right\rangle$ is specific to higher-spin systems, because nontrivial quadrupolar degrees of freedom are absent for spin-$\frac{1}{2}$. The phase diagram presented here is relevant to two triangular-lattice nickelate compounds discussed in the Introduction. Na$_2$BaNi(PO$_4$)$_2$, which hosts a relatively large single-ion anisotropy of $D_{z} \approx 4$, lies in the large-$D_z$ regime of the phase diagram. The dynamical spin structure factors obtained for various $h_{z}$, presented in Sec.~\ref{subsec:SS_Dz4:Dynamical}, provide direct theoretical predictions for comparison with the INS spectra of this material. By contrast, K$_2$Ni(SeO$_3$)$_2$ exhibits a much weaker single-ion anisotropy, placing it close to the $D_z = 0$ limit. While K$_2$Ni(SeO$_3$)$_2$ shows weak easy-plane exchange anisotropy~\cite{li2023k2ni,kim2026spin}, the Zeeman energy becomes increasingly important at high magnetic fields. Therefore, the regime near $D_z = 0$ can provide possible field-induced phases relevant to K$_2$Ni(SeO$_3$)$_2$, whose characteristic signatures may be tested through magnetization measurements and INS.

\section{spin supersolids at $D_z =0$}
\label{sec:SS_Dz0}

\subsection{spin supersolid orders}
\label{subsec:SS_Dz0:SS}
We study the quantum phase transitions at $D_z = 0$ as a function of the magnetic field $h_z$. To identify the nature of each phase, we examine four order parameters: the transverse dipolar order at the $\mathbf{K}$ points $\left\langle m^2_{xy}(\mathbf{K}) \right\rangle$; the longitudinal dipolar order at the $\mathbf{K}$ points $\left\langle m^2_z(\mathbf{K}) \right\rangle$; and the quadrupolar orders at both $\mathbf{\Gamma}$ points $\left\langle Q^2_\perp(\mathbf{\Gamma}) \right\rangle$ and $\mathbf{K}$ points $\left\langle Q^2_\perp(\mathbf{K}) \right\rangle$. The locations of $\mathbf{K}$ points and $\mathbf{\Gamma}$ points in the Brillouin zone are illustrated in Figs.~\ref{Fig1_phase_diagram} (c) and (d), respectively.

\begin{figure*}
\centering
\includegraphics[width=0.95\linewidth]{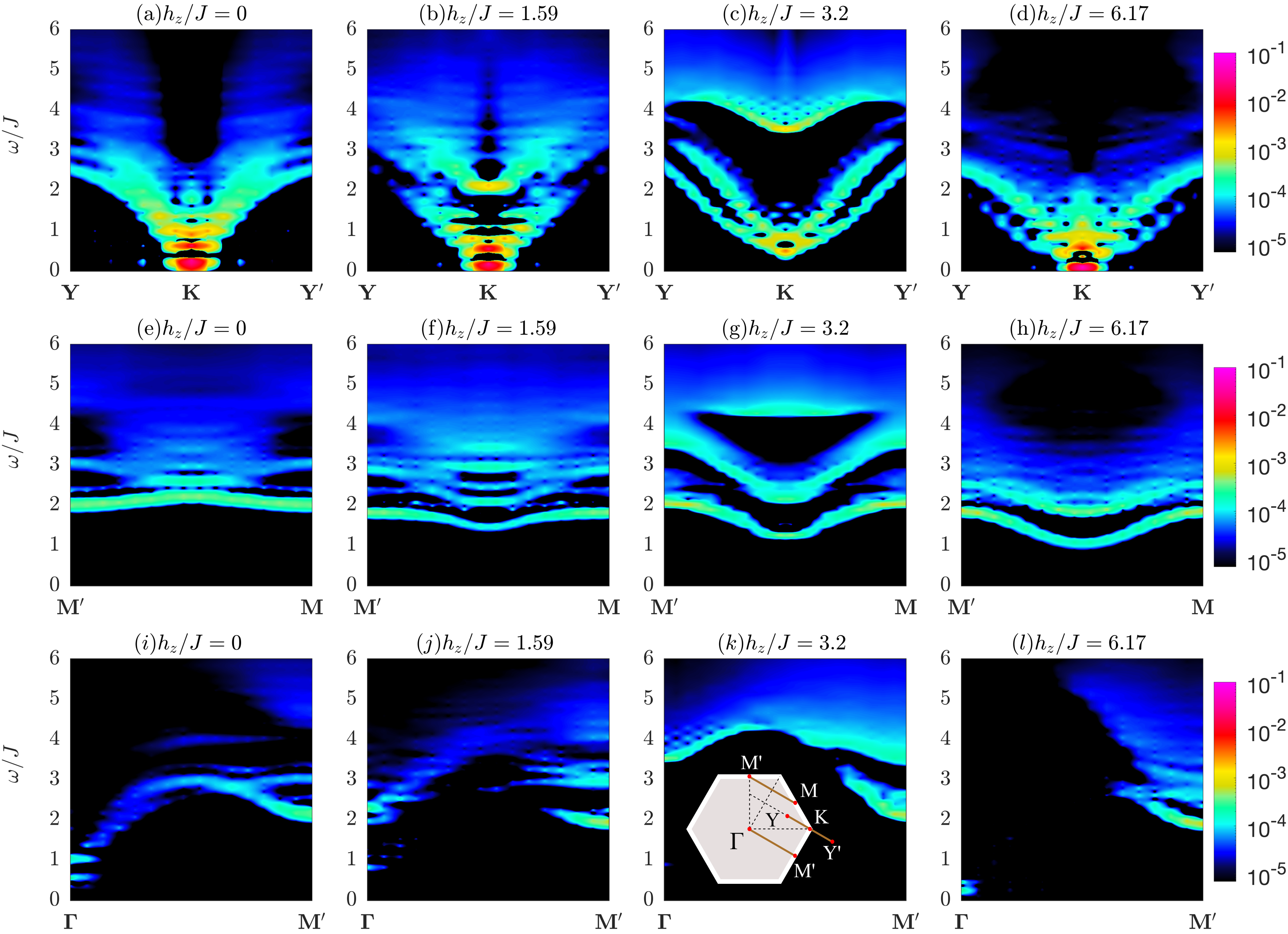}
\caption{The transverse component of the dynamical spin structure factor $\chi^{xy}(\mathbf{k},\omega)$ for various $h_{z}$ at $D_{z}=0$. Panels (a), (e) and (i) are obtained in the dipolar YSS at zero field. Panels (b), (f) and (j) are obtained in the dipolar YSS at finite field of $h_{z}=1.59$. Panels (c), (g) and (k) are obtained in the UUD phase. Panels (d), (h) and (l) are obtained in the dipolar VSS phase. Results in the SS and UUD phases are calculated on the lattice of $N = 36 \times 6$ and $N = 48 \times 6$, respectively.}
\label{Fig_Dz0_spin_xy_stru}
\end{figure*}

We first identify the ground state at high magnetic fields, using $h_z = 6.17$ as a representative example. As shown in Fig.~\ref{Fig_Dz0} (a), all four order parameters exhibit almost no dependence on the system length $L_x$, indicating very little finite-size effects along the $x$ direction. We therefore perform finite-size scaling with respect to the system width $L_y$, as shown in Fig.~\ref{Fig_Dz0} (b). All four order parameters remain finite in the thermodynamic limit, with the dipolar order $\left\langle m^2_{xy}(\mathbf{K}) \right\rangle$ being much larger than the quadrupolar orders $\left\langle Q^2_\perp(\mathbf{\Gamma}) \right\rangle$ and $\left\langle Q^2_\perp(\mathbf{K}) \right\rangle$. Furthermore, the real space distribution of $\left\langle S_{i}^{z}\right\rangle$ exhibits a `V'-type spin configuration, as shown in Appendix~\ref{Apendix_real_space}. The dominant transverse dipolar order coexisting with the longitudinal translational symmetry breaking with `V'-type spin configuration is consistent with the dipolar VSS.

We then turn to the states at low magnetic fields. We find again that the physical quantities show very little dependence on the system length $L_{x}$. Figure~\ref{Fig_Dz0} (c) shows the finite-size extrapolation with respect to $L_y$ at $h_z = 1.59$. In contrast to the high-field case, the quadrupolar orders $\left\langle Q^2_\perp(\mathbf{\Gamma}) \right\rangle$ and $\left\langle Q^2_\perp(\mathbf{K}) \right\rangle$ extrapolate to zero in the infinite $L_y$ limit, while $\left\langle m^2_{xy}(\mathbf{K}) \right\rangle$ and $\left\langle m^2_z(\mathbf{K}) \right\rangle$ remain finite. In addition, `Y'-type spin configurations are found in real space distribution of $\left\langle S_{i}^{z}\right\rangle$. The coexistence of finite transverse dipolar order and longitudinal translational symmetry breaking with `Y'-type spin configurations, identifies the phase as the dipolar YSS. At lower magnetic fields down to zero, a weak but finite $\left\langle Q^2_\perp(\mathbf{\Gamma}) \right\rangle$ survives in the thermodynamic limit within part of the dipolar YSS phase. In the zero field limit at $D_z =0$, the Hamiltonian only consists of dipolar interactions. Thus, the weak quadrupolar order may be induced by the primary dipolar order and the phase is named as dipolar SS. Further details of the finite-size scaling are provided in Appendix~\ref{Apendix_finite_size}.

To map out the phase diagram and locate the phase boundaries, we calculate the order parameters for various $h_z$ with a fixed lattice size. As shown in Fig.~\ref{Fig_Dz0} (d), $\left\langle m^2_z(\mathbf{K}) \right\rangle$ increases as $h_{z}$ increases from 0, and reaches a maximum near $h_z \approx 2.82$. On the other hand, $\left\langle m^2_{xy}(\mathbf{K}) \right\rangle$ is largely suppressed and reaches a minimum at the same $h_{z}$, signaling a phase transition from the dipolar YSS phase to the UUD phase, which is characterized by a three-sublattice longitudinal magnetic order and a $1/3$ magnetization plateau. A second transition from the UUD phase to the dipolar VSS is identified near $h_z \approx 4.1$, where $\left\langle m^2_{xy}(\mathbf{K}) \right\rangle$ increases and $\left\langle m^2_z(\mathbf{K}) \right\rangle$ decreases. The smooth evolution of $\left\langle m^2_{xy}(\mathbf{K}) \right\rangle$ across the phase boundary is consistent with a continuous transition. In contrast, for larger $D_{z} > 1.5$, a sudden jump in $\left\langle m^2_{xy}(\mathbf{K}) \right\rangle$ near the transition point is identified to suggest a first-order transition from the UUD phase to the dipolar VSS; detailed results are presented in Appendix~\ref{Apendix_fixed_lattice}. Finally, a transition into the fully polarized phase occurs near $h_z \approx 9.34$.

As shown in Fig.~\ref{Fig_Dz0} (e), the quadrupolar moments remain weak in the low-field YSS regime, while $\left\langle Q^2_\perp(\mathbf{\Gamma}) \right\rangle$ develops a moderate peak in the high-field VSS regime. Based on the finite-size extrapolations of $\left\langle Q^2_\perp(\mathbf{\Gamma}) \right\rangle$ and $\left\langle Q^2_\perp(\mathbf{K}) \right\rangle$ which are presented in Appendix~\ref{Apendix_finite_size}, we map out the shaded regimes in the dipolar VSS and YSS phases where both $\left\langle Q^2_\perp(\mathbf{\Gamma}) \right\rangle$ and $\left\langle Q^2_\perp(\mathbf{K}) \right\rangle$ vanish in the thermodynamic limit.

\subsection{Spin superfluid stiffness}
\label{subsec:SS_Dz0:stiff}
The superfluid stiffness is examined through the energy difference by a twist angle, which is $\Delta E_0(\pi/2)$ as given in Eq.~\ref{eq_stiffness}. $\theta =\pi/2$ is chosen to capture contributions from both quadrupolar and dipolar orders. As shown in Fig.~\ref{Fig_Dz0} (f), the evolution of $\Delta E_0(\pi/2)$ as a function of $h_z$ is consistent with the quantum phase diagram. It remains finite in the dipolar YSS and VSS regimes, but vanishes in the UUD and polarized phases, with the corresponding phase boundaries consistent with those determined from the order parameters.

\subsection{Dynamical spin structure factors}
\label{subsec:SS_Dz0:Dynamical}

\begin{figure*}
\centering
\includegraphics[width=0.75\linewidth]{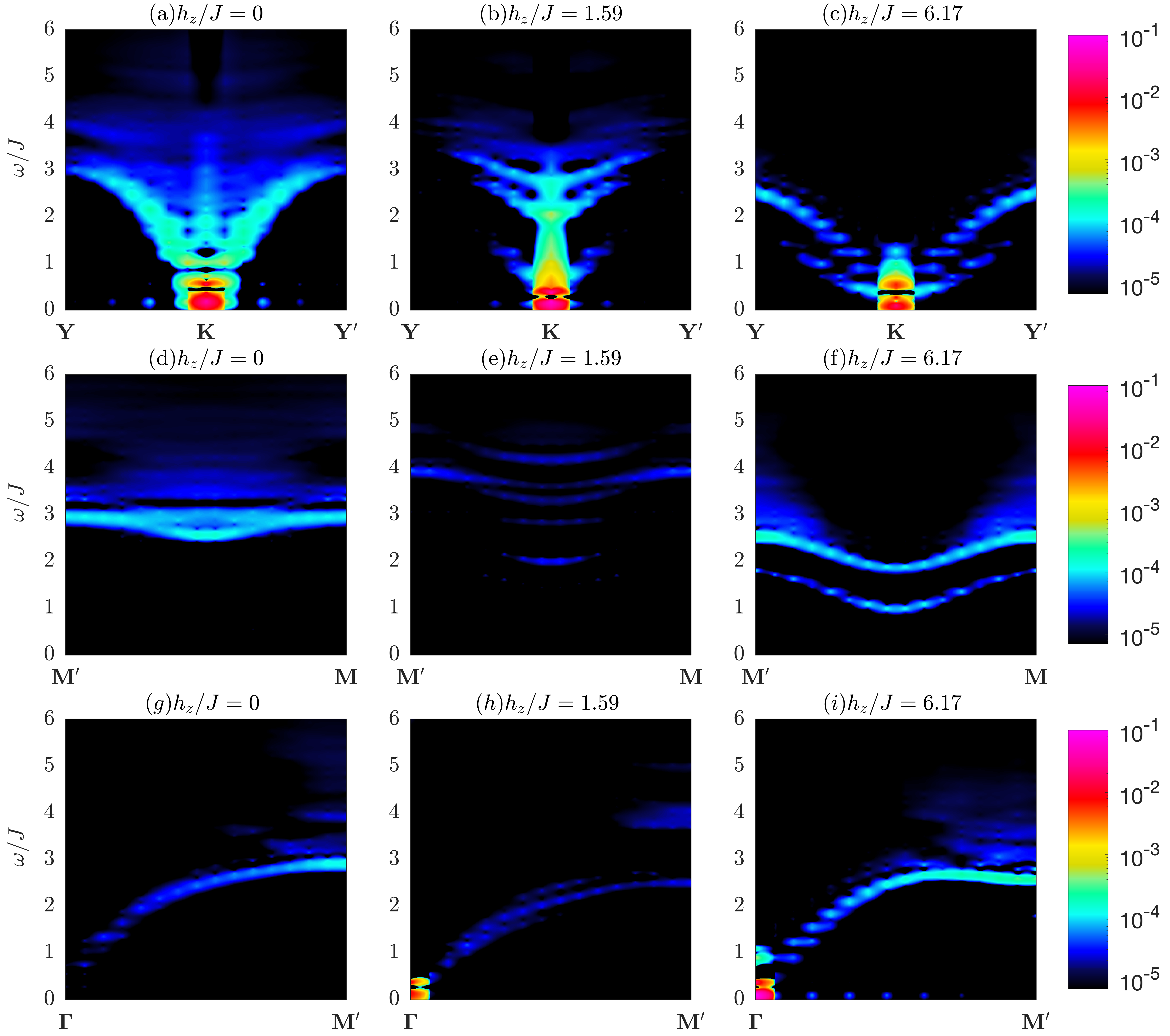}
\caption{Same as Fig.~\ref{Fig_Dz0_spin_xy_stru}, but for the longitudinal component of the dynamical spin structure factor $\chi^{z}(\mathbf{k},\omega)$. In the UUD state, the dynamical spin structure factor is dominated by $\chi^{xy}(\mathbf{k},\omega)$, thus $\chi^{z}(\mathbf{k},\omega)$ is not shown here.}
\label{Fig_Dz0_spin_z_stru}
\end{figure*}

The transverse $\chi^{xy}(\mathbf{k},\omega)$ and the longitudinal $\chi^{z}(\mathbf{k},\omega)$ dynamical spin structure factor are defined as 
\begin{align}
\label{eq_dynamical_spin_factor}
\chi^{xy}(\mathbf{k},\omega)=&\frac{1}{3N_{bulk}}\sum_{i_0,j}\int_{0}^{\tau_{tot}}d\tau e^{i\omega \tau-\eta \tau} e^{-i\mathbf{k}\cdot (\mathbf{r}_{j}-\mathbf{r}_{i_0})} \times \nonumber \\
&\frac{1}{2}\langle S_{i_0}^{+}(\tau )S_{j}^{-}+H.c.\rangle. \nonumber \\
\chi^{z}(\mathbf{k},\omega)=&\frac{1}{3N_{bulk}}\sum_{i_0,j}\int_{0}^{\tau_{tot}}d\tau e^{i\omega \tau-\eta \tau} e^{-i\mathbf{k}\cdot (\mathbf{r}_{j}-\mathbf{r}_{i_0})} \times \\
&\langle S_{i_0}^{z}(\tau )S_{j}^{z}\rangle . \nonumber
\end{align}
To avoid the boundary effect, the summation of $j$ is chosen to be over the $N_{bulk}=\frac{2}{3}L_{x}\times L_{y}$ sites at the center of the lattice. $i_0$ is summed over a three-site unit cell in the center. Due to the finite time that can be numerically simulated, a smearing factor of $e^{-\eta \tau}$ is applied in the Fourier transform from time domain to $\omega$ domain, which is followed by the Fourier transform from real space to momentum space. Here $\eta =1/\tau _{tot}$ and $\tau _{tot}$ is the total simulation time.

We investigate the dynamical properties at $D_z = 0$ by examining the transverse component and the longitudinal component of the dynamical spin structure factor at four representative magnetic fields spanning the dipolar YSS, UUD, and dipolar VSS phases. The results of $\chi^{xy}(\mathbf{k}, \omega)$ and $\chi^{z}(\mathbf{k}, \omega)$ are shown in Fig.~\ref{Fig_Dz0_spin_xy_stru} and Fig.~\ref{Fig_Dz0_spin_z_stru}, respectively. In both figures, the top, middle, and bottom rows show the spectra along the momentum paths $\mathbf{Y}$-$\mathbf{K}$-$\mathbf{Y'}$, $\mathbf{M'}$-$\mathbf{M}$, and $\mathbf{ \Gamma }$-$\mathbf{M'}$, respectively. The corresponding momentum paths are illustrated in the inset of Fig.~\ref{Fig_Dz0_spin_xy_stru} (k).

At zero field, the ground state is the dipolar YSS. As shown in Fig.~\ref{Fig_Dz0_spin_xy_stru} (a), $\chi^{xy}(\mathbf{k}, \omega)$ along the $\mathbf{Y}$-$\mathbf{K}$-$\mathbf{Y'}$ path exhibits a gapless Goldstone mode at the $\mathbf{K}$ point, which is a hallmark of the spontaneously broken $U(1)$ symmetry associated with the supersolid order. The discrete spectral weights in the spectrum may be due to the finite-size effect. Similarly, the gapless Goldstone mode can be found in $\chi^{z}(\mathbf{k}, \omega)$, as shown in Fig.~\ref{Fig_Dz0_spin_z_stru} (a). Along the $\mathbf{M'}$-$\mathbf{M}$ path, Fig.~\ref{Fig_Dz0_spin_xy_stru} (e) reveals a roton-like minimum in the lowest dispersive mode near the $\mathbf{M}$ points of $\chi^{xy}(\mathbf{k}, \omega)$, similar to the roton-like feature previously identified in the dynamical spin structure factor of spin-$\frac{1}{2}$ triangular-lattice supersolids~\cite{huang2026dissipationless}. However, such roton-like minimum is not observed in $\chi^{z}(\mathbf{k}, \omega)$, as shown in Fig.~\ref{Fig_Dz0_spin_z_stru} (d). Interestingly, finite spectral weight extends to higher energies near the $\mathbf{M}$ points, which is consistent with the multiparticle continuum proposed in a recent study~\cite{kim2026spin}. Comparing Fig.~\ref{Fig_Dz0_spin_xy_stru} (e) and Fig.~\ref{Fig_Dz0_spin_z_stru} (d), the spectral weight at $\omega /J> 4$ arises predominantly from the transverse component $\chi^{xy}(\mathbf{k}, \omega)$. The high-energy spectral weight is further enhanced as the magnetic field is increased to $1.59$, as shown in Fig.~\ref{Fig_Dz0_spin_xy_stru} (f). Furthermore, as shown in Figs.~\ref{Fig_Dz0_spin_xy_stru} (i) and Fig.~\ref{Fig_Dz0_spin_z_stru} (g), dispersive modes in both $\chi^{xy}(\mathbf{k}, \omega)$ and $\chi^{z}(\mathbf{k}, \omega)$ descend to low energies near $\mathbf{\Gamma}$ points with much smaller intensity. Similar low-energy spectral features near $\mathbf{\Gamma}$ points have also been observed in the SS phase of spin-$\frac{1}{2}$ triangular-lattice systems~\cite{chi2024dynamical}.

\begin{figure*}
\centering
\includegraphics[width=0.95\linewidth]{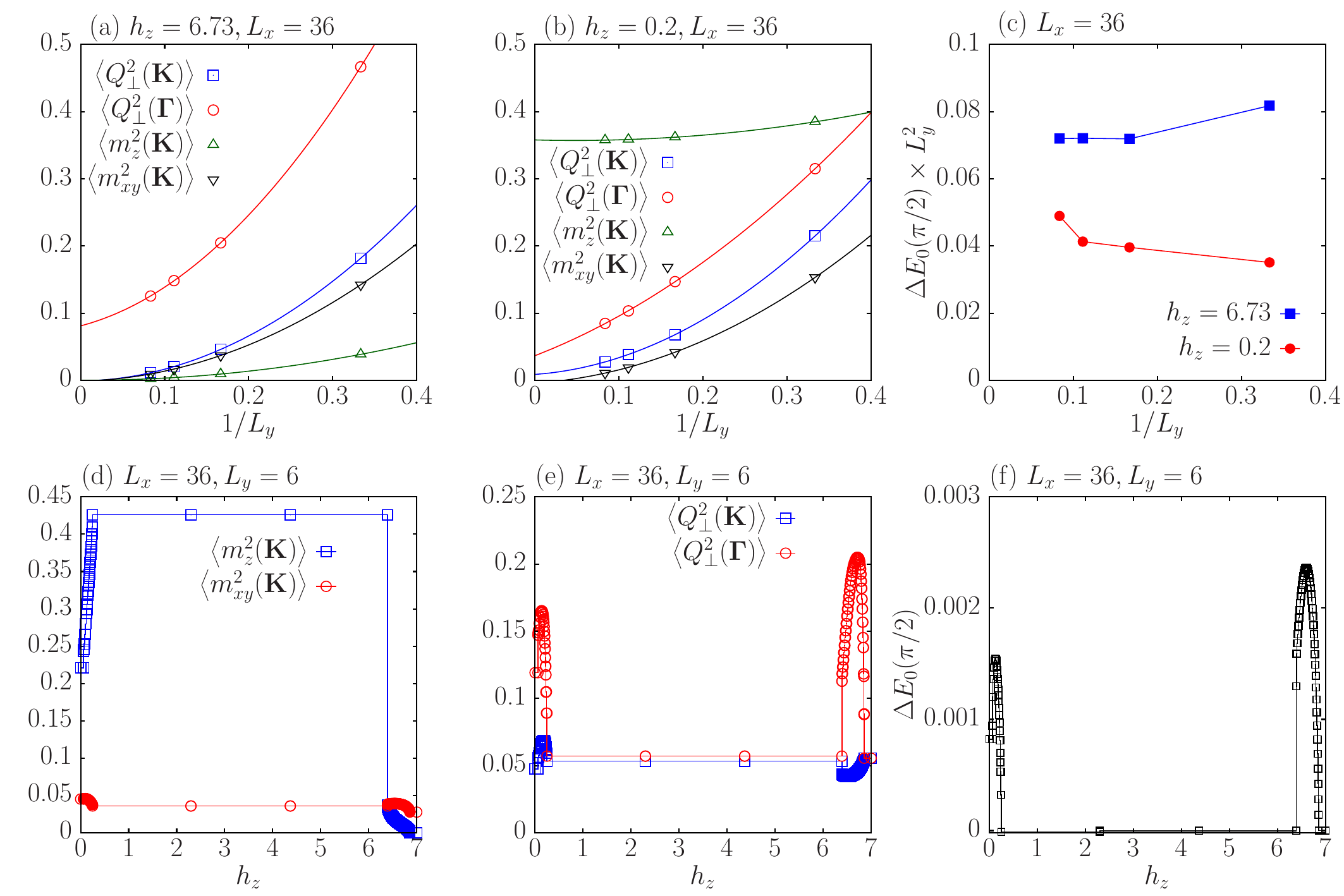}
\caption{Panels (a) and (b) show the finite-size scaling in $L_{y}$ for various orders at $h_{z}=6.73$ and $h_{z}=0.2$, respectively. Panel (c) shows the finite-size scaling of the superfluid stiffness. Panels (d), (e), and (f) show the evolution of various orders and superfluid stiffness for different magnetic fields $h_{z}$ on a fixed lattice size. All results are obtained at $D_{z}=4$.}
\label{Fig_Dz4_order}
\end{figure*}

We then consider the dipolar YSS at $h_z = 1.59$. As shown in Fig.~\ref{Fig_Dz0_spin_xy_stru} (b), $\chi^{xy}(\mathbf{k}, \omega)$ along the $\mathbf{Y}$-$\mathbf{K}$-$\mathbf{Y'}$ path exhibits a gapless Goldstone mode near the $\mathbf{K}$ points, confirming the persistence of spontaneously broken U(1) symmetry at this finite field. $\chi^{z}(\mathbf{k}, \omega)$ also shows the gapless mode near the $\mathbf{K}$ points with strongly dispersive spectral features, as shown in Fig.~\ref{Fig_Dz0_spin_z_stru} (b). In contrast to the zero-field spectrum discussed above, an additional higher-energy magnon branch emerges above the Goldstone mode in both $\chi^{xy}(\mathbf{k}, \omega)$ and $\chi^{z}(\mathbf{k}, \omega)$. Furthermore, the roton-like minimum near the $\mathbf{M}$ points, which was clearly visible at $h_z = 0$, is no longer observed at $h_z = 1.59$, as shown in Fig.~\ref{Fig_Dz0_spin_xy_stru} (f). A similar field-induced suppression of the roton-like minimum has been found in the spin-$\frac{1}{2}$ systems~\cite{huang2026dissipationless}. Near $\mathbf{\Gamma}$ points, the dispersive mode in $\chi^{xy}(\mathbf{k}, \omega)$ identified at zero field is pushed to higher energies at $h_z = 1.59$ [Fig.~\ref{Fig_Dz0_spin_xy_stru} (j)], while it remains almost unchanged for $\chi^{z}(\mathbf{k}, \omega)$ [Fig.~\ref{Fig_Dz0_spin_z_stru} (h)].

At the intermediate field $h_z = 3.2$, the ground state is the UUD state, characterized by a three-sublattice order along the $z$ direction and a $1/3$ magnetization plateau. Accordingly, the dominant dynamical spectral weight appears in $\chi^{xy}(\mathbf{k}, \omega)$. As shown in Figs.~\ref{Fig_Dz0_spin_xy_stru} (c) and (g), the dynamical spectrum is qualitatively different from the SS phases. It consists of three well-defined dispersive bands with a finite energy gap. The gapped spectrum and the absence of any gapless Goldstone mode are consistent with the absence of spontaneous U(1) symmetry breaking in the UUD phase.

At the higher field $h_z = 6.17$, the ground state is the dipolar VSS. Both $\chi^{xy}(\mathbf{k}, \omega)$ [Fig.~\ref{Fig_Dz0_spin_xy_stru} (d)] and $\chi^{z}(\mathbf{k}, \omega)$ [Fig.~\ref{Fig_Dz0_spin_z_stru} (c)] again exhibit the gapless Goldstone mode at the $\mathbf{K}$ point with a dispersive excitations qualitatively similar to that at $h_z = 0$, confirming the supersolid nature of this phase. However, the high-field spectra differ from the zero-field spectra in two notable respects. First, the roton-like minimum near the $\mathbf{M}$ points is absent in $\chi^{xy}(\mathbf{k}, \omega)$, as shown in Fig.~\ref{Fig_Dz0_spin_xy_stru} (h). Second, the low-energy mode near $\mathbf{\Gamma}$ observed at zero field is absent in $\chi^{xy}(\mathbf{k}, \omega)$ [Fig.~\ref{Fig_Dz0_spin_xy_stru} (l)], while the low-energy mode persists in $\chi^{z}(\mathbf{k}, \omega)$ [Fig.~\ref{Fig_Dz0_spin_z_stru} (i)]. These distinct spectral features provide spectroscopic signatures of the dipolar SS that could, in principle, be resolved by spin-polarized INS experiments.



\section{spin supersolids at $D_z= 4$}
\label{sec:SS_Dz4}
\subsection{spin supersolid orders}
\label{subsec:SS_Dz4:SS}
At large single-ion anisotropy, $D_z = 4$, because of the high energy cost of occupying the $|S^z =0\rangle$ states, the ground state without magnetic fields favors a quadrupolar order, in which the transverse quadrupolar moment spontanously breaks the spin U(1) symmetry with a finite spin superfluid stiffness, but the transverse dipolar order remains absent. This is also known as the spin-nematic state with `hidden order'~\cite{sheng2025bose}, as the spin rotational symmetry is spontaneously broken without any dipolar magnetic order. The state is labeled as the quadrupolar SS in Fig.~\ref{Fig1_phase_diagram} (a) which continuously connected to finite fields. As the magnetic field increases, the competition among the Zeeman coupling, exchange interactions, and single-ion anisotropy drives the system into the UUD phase. Upon further increasing the field, the system transits into a quadrupolar spin superfluid, in which uniform transverse quadrupolar order develops at $\mathbf{\Gamma} $ points without any longitudinal translational symmetry breaking. The system eventually reaches the fully polarized phase at sufficiently high fields.

\begin{figure*}
\centering
\includegraphics[width=0.95\linewidth]{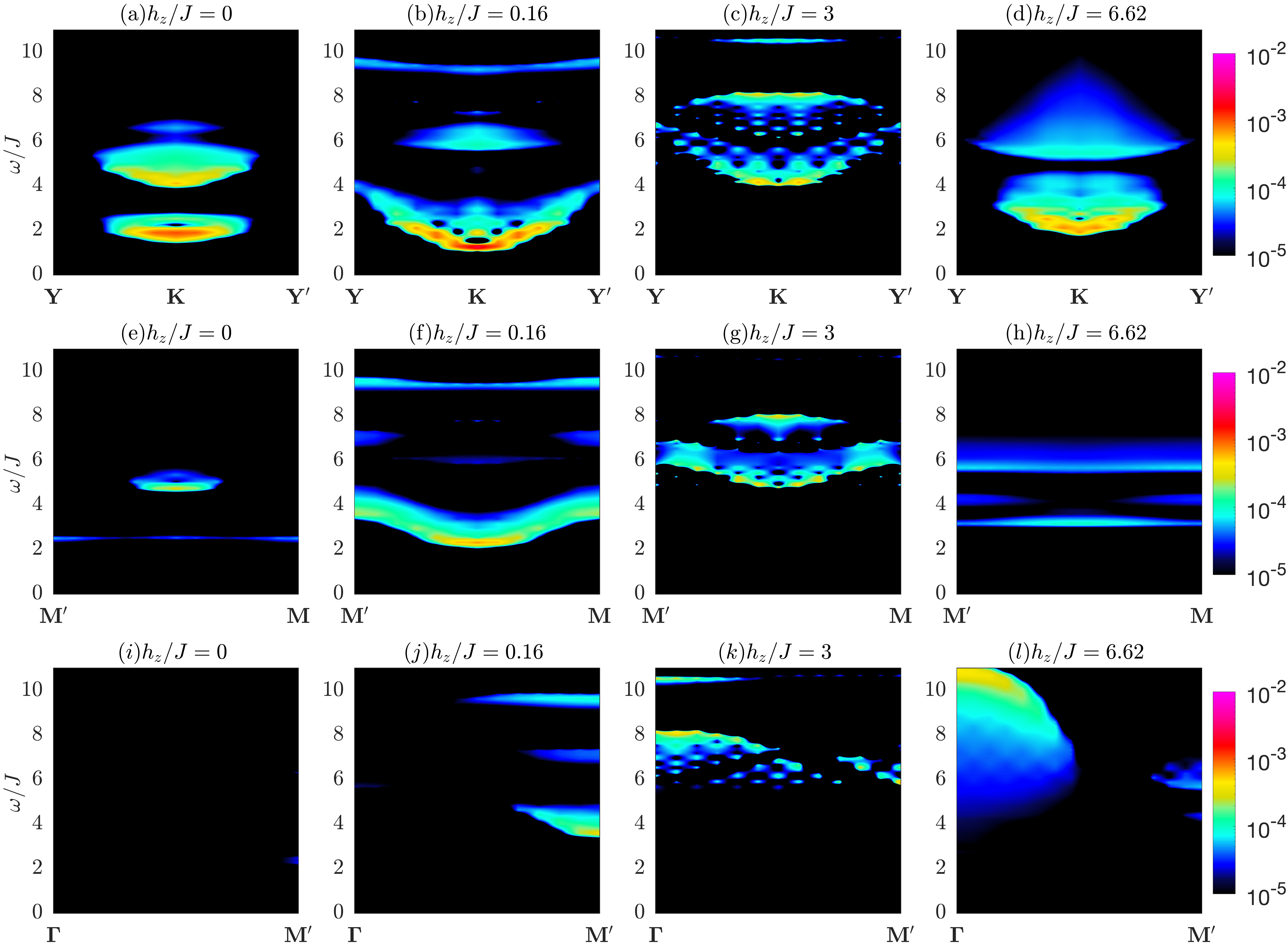}
\caption{The transverse component of the dynamical spin structure factor $\chi^{xy}(\mathbf{k},\omega)$ for various $h_{z}$ at $D_{z}=4$. Panels (a), (b), (e), (f), (i) and (j) are obtained in the quadrupolar SS. Panels (c), (g) and (k) are obtained in the UUD phase. Panels (d), (h) and (l) are obtained in the quadrupolar spin superfluid. All results are calculated on the lattice of $N = 36 \times 6$.}
\label{Fig_Dz4_spin_xy_stru}
\end{figure*}

We determine the dipolar and quadrupolar orders by calculating the order parameters $\left\langle m^2_{xy}(\mathbf{K}) \right\rangle$, $\left\langle m^2_z(\mathbf{K}) \right\rangle$, $\left\langle Q^2_\perp(\mathbf{K}) \right\rangle$, and $\left\langle Q^2_\perp(\mathbf{\Gamma}) \right\rangle$ with finite-size scaling, and further characterize the phases through the superfluid stiffness $\Delta E_0(\pi/2)$.

We first identify the ground state at the high field $h_z = 6.73$. Similar to the $D_{z}=0$ case, the states show little $L_{x}$ dependence. As shown in Fig.~\ref{Fig_Dz4_order} (a), the finite-size scaling of $L_y$ reveals that $\left\langle m^2_{xy}(\mathbf{K}) \right\rangle$, $\left\langle m^2_z(\mathbf{K}) \right\rangle$, and $\left\langle Q^2_\perp(\mathbf{K}) \right\rangle$ all extrapolate to zero in the thermodynamic limit, while $\left\langle Q^2_\perp(\mathbf{\Gamma}) \right\rangle$ remains finite. The absence of $\left\langle m^2_{xy}(\mathbf{K}) \right\rangle$ and $\left\langle m^2_z(\mathbf{K}) \right\rangle$, combined with the uniform transverse quadrupolar order at $\mathbf{\Gamma}$ points, identifies this phase as a quadrupolar spin superfluid.

At the low field $h_z = 0.2$, as shown in Fig.~\ref{Fig_Dz4_order} (b), the finite-size scaling yields finite values of both $\left\langle Q^2_\perp(\mathbf{\Gamma}) \right\rangle$ and $\left\langle Q^2_\perp(\mathbf{K}) \right\rangle$, indicating a possible ``multi-Q'' feature in the quadrupolar structure factor. Despite the presence of these multiple ordering wave vectors, no finite scalar spin chirality is detected in the ground states. Furthermore, $\left\langle m^2_z(\mathbf{K}) \right\rangle$ extrapolates to a finite value while $\left\langle m^2_{xy}(\mathbf{K}) \right\rangle$ extrapolates to zero in the thermodynamic limit, confirming the absence of transverse dipolar order. This combination of finite $\left\langle Q^2_\perp(\mathbf{K}) \right\rangle$, $\left\langle Q^2_\perp(\mathbf{\Gamma}) \right\rangle$, and $\left\langle m^2_z(\mathbf{K}) \right\rangle$, together with vanishing $\left\langle m^2_{xy}(\mathbf{K}) \right\rangle$, identifies the low-field ground state as the quadrupolar SS.

To map out the field dependence and locate the phase boundaries between these phases, we scan the order parameters as a function of $h_z$ on a fixed cylinder of $N = 36 \times 6$. The dipolar order parameters $\left\langle m^2_{xy}(\mathbf{K}) \right\rangle$ and $\left\langle m^2_z(\mathbf{K}) \right\rangle$ are shown in Fig.~\ref{Fig_Dz4_order} (d), and the quadrupolar order parameters $\left\langle Q^2_\perp(\mathbf{\Gamma}) \right\rangle$ and $\left\langle Q^2_\perp(\mathbf{K}) \right\rangle$ are shown in Fig.~\ref{Fig_Dz4_order} (e). At low fields, $\left\langle m^2_z(\mathbf{K}) \right\rangle$, $\left\langle Q^2_\perp(\mathbf{\Gamma}) \right\rangle$, and $\left\langle Q^2_\perp(\mathbf{K}) \right\rangle$ remain finite but $\left\langle m^2_{xy}(\mathbf{K}) \right\rangle$ is suppressed, which is consistent with the quadrupolar SS. Near $h_z \approx 0.25$, $\left\langle Q^2_\perp(\mathbf{\Gamma}) \right\rangle$ and $\left\langle Q^2_\perp(\mathbf{K}) \right\rangle$ reach minima while $\left\langle m^2_z(\mathbf{K}) \right\rangle$ reaches a maximum, indicating a phase transition into the UUD phase. The UUD phase is characterized by the maximum value of $\left\langle m^2_z(\mathbf{K}) \right\rangle$ in the intermediate regime, because the spins align in the $z$ direction. A second phase transition into the quadrupolar spin superfluid is identified near $h_z \approx 6.4$, where $\left\langle Q^2_\perp(\mathbf{\Gamma}) \right\rangle$ shows a sudden rise and $\left\langle m^2_z(\mathbf{K}) \right\rangle$ shows a sudden drop, which is consistent with a first-order transition~\cite{sheng2025possible,huang2025universal}. Furthermore, a phase transition into the fully polarized state occurs near $h_z \approx 6.86$ where $\left\langle Q^2_\perp(\mathbf{\Gamma}) \right\rangle$ reaches minimum.

\subsection{Superfluid stiffness}
\label{subsec:SS_Dz4:stiff}
To further establish the spin superfluidity nature of the phases, we calculate the spin superfluid stiffness for various system widths $L_y$. Because the twisted phase $\theta$ at the $y$ boundary is equivalent to $\theta /L_{y}$ added to every spin-flip term in the $y$ direction through a gauge transformation~\cite{huang2026dissipationless}, we calculate $\Delta E_{0}(\pi / 2)\times L_{y}^{2}$ to directly compare results obtained for systems with different $L_{y}$. As shown in Fig.~\ref{Fig_Dz4_order} (c), $\Delta E_0(\pi/2) \times L_y^2$ remains robust with increasing $L_y$, indicating a finite superfluid stiffness in both the quadrupolar SS and the quadrupolar spin superfluid. The finite superfluid stiffness in the quadrupolar SS and spin superfluid phases originates from the spontaneous spin U(1) symmetry breaking by the transverse quadrupolar order, in contrast to the conventional spin superfluidity associated with transverse dipolar magnetic order~\cite{yuan2018experimental}.

The superfluid stiffness for various $h_z$ provides a consistent characterization of the quantum phase transitions, as shown in Fig.~\ref{Fig_Dz4_order} (f). Although $\Delta E_0(\pi/2)$ at $D_{z}=4$ is substantially smaller than that at $D_{z}=0$, it remains finite in the quadrupolar SS at low fields and quadrupolar spin superfluid at high fields, while it vanishes within the numerical accuracy in both the UUD and fully polarized phases. The phase boundaries indicated by the superfluid stiffness are consistent with those identified from the order parameters, providing independent numerical evidence to identify the phases at $D_z = 4$.

\begin{figure*}
\centering
\includegraphics[width=0.75\linewidth]{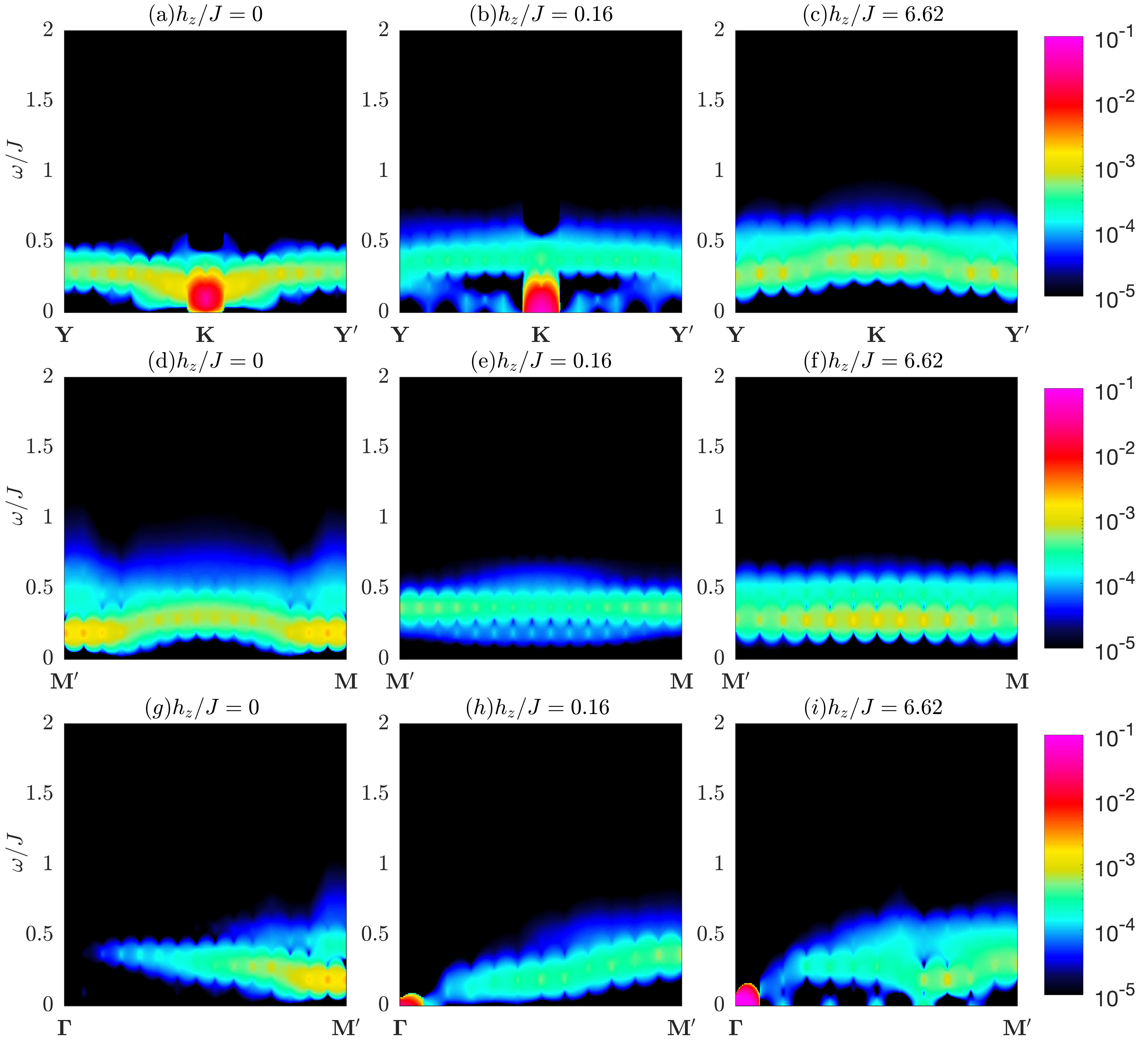}
\caption{Same as Fig.~\ref{Fig_Dz4_spin_xy_stru}, but for the longitudinal component of the dynamical spin structure factor $\chi^{z}(\mathbf{k},\omega)$. The $\chi^{z}(\mathbf{k},\omega)$ for the UUD phase is not shown because the spectrum is dominated by the transverse dynamical spin structure factor.}
\label{Fig_Dz4_spin_z_stru}
\end{figure*}


\subsection{Dynamical spin structure factor}
\label{subsec:SS_Dz4:Dynamical}
The dynamical spin structure factor is directly accessible in INS experiments, and it provides valuable insights into the nature of the SS states. It is useful to separately examine the transverse dynamical spin structure factor $\chi^{xy}(\mathbf{k}, \omega)$ and the longitudinal dynamical spin structure factor $\chi^z(\mathbf{k}, \omega)$, because their spectral weights are well separated in energy at $D_z = 4$. The low-energy excitations predominantly appear in $\chi^z(\mathbf{k}, \omega)$, including the gapless modes, while the spectral weight of $\chi^{xy}(\mathbf{k}, \omega)$ is concentrated at higher energies. This is qualitatively different from that at $D_{z}=0$, where both $\chi^{xy}(\mathbf{k}, \omega)$ and $\chi^z(\mathbf{k}, \omega)$ exhibit gapless Goldstone modes in the SS states. Results of $\chi^{xy}(\mathbf{k}, \omega)$ [Fig.~\ref{Fig_Dz4_spin_xy_stru}] and $\chi^z(\mathbf{k}, \omega)$ [Fig.~\ref{Fig_Dz4_spin_z_stru}] are shown along different momentum paths of $\mathbf{Y}$-$\mathbf{K}$-$\mathbf{Y'}$, $\mathbf{M'}$-$\mathbf{M}$, and $\mathbf{ \Gamma }$-$\mathbf{M'}$ at the top, middle, and bottom rows, respectively.

We first discuss the quadrupolar SS state at zero and low fields. Because this state lacks transverse dipolar order, $\chi^{xy}(\mathbf{k}, \omega)$ shows spectral weight predominantly at higher energies, with dispersive excitations near the $\mathbf{K}$ points and $\mathbf{M}$ points, as shown in Figs.~\ref{Fig_Dz4_spin_xy_stru} (a), (e), (i) for $h_z = 0$, and Figs.~\ref{Fig_Dz4_spin_xy_stru} (b), (f), (j) for $h_z =0.16$. By contrast, $\chi^z(\mathbf{k}, \omega)$ exhibits a gapless excitation at the $\mathbf{K}$ points at zero field [Fig.~\ref{Fig_Dz4_spin_z_stru} (a)], which is consistent with the gapless Goldstone excitation arising from the spontaneously broken U(1) symmetry by the quadrupolar order. As the field increases, the low-energy spectral weight of $\chi^z(\mathbf{k}, \omega)$ becomes increasingly concentrated at $\mathbf{\Gamma}$ points, as shown in Figs.~\ref{Fig_Dz4_spin_z_stru} (g) and (h). In addition to the gapless excitation, Fig.~\ref{Fig_Dz4_spin_z_stru} (b) shows a gapped mode near the $\mathbf{K}$ points, which is consistent with previous results studying the effective two-level model $H_{\textit{eff}}$~\cite{sheng2025possible}. The large intensity near zero energy originates from the static contribution of $\left \langle S^{z}_{i} \right \rangle$ which exhibits a three-sublattice spatial modulation pinned by the open boundary. In addition, the roton-like minimum is observed near the $\mathbf{M}$ points at zero field [Fig.~\ref{Fig_Dz4_spin_z_stru} (d)], but becomes less pronounced when the magnetic field is increased to $h_{z}=0.16$ [Fig.~\ref{Fig_Dz4_spin_z_stru} (e)].

In the UUD phase, $\chi^{xy}(\mathbf{k}, \omega)$ exhibits several dispersive bands, with finite spectral weight between the bands, as shown in Figs.~\ref{Fig_Dz4_spin_xy_stru} (c), (g), and (k). This interband spectral weight is not observed in the dynamical spin structure factor of the UUD phase for the spin-$\frac{1}{2}$ system~\cite{huang2026dissipationless}, where the spectrum consists of well-separated magnon bands. Furthermore, nearly flat high-energy excitations appear above the dispersive bands, which is consistent with previous studies that suggest a three-magnon
continuum~\cite{sheng2025possible}. These distinctions show a richer excitation spectrum in the spin-1 systems, potentially reflecting multipolar fluctuations and multi-magnon scattering channels that are usually suppressed in spin-$\frac{1}{2}$ systems. In contrast, $\chi^z(\mathbf{k}, \omega)$ shows almost vanishing spectral weights throughout the Brillouin zone in the UUD phase. This behavior is consistent with the predominantly longitudinal order in the UUD phase, so that the excitations primarily involve transverse fluctuations.

In the quadrupolar spin superfluid at higher fields, $\chi^{xy}(\mathbf{k}, \omega)$ resembles the spectrum obtained in the low-field quadrupolar SS, as shown in Fig.~\ref{Fig_Dz4_spin_xy_stru} (d). The low-energy structure of $\chi^z(\mathbf{k}, \omega)$, however, is qualitatively different. As shown in Figs.~\ref{Fig_Dz4_spin_z_stru} (c) and (i), a small gap appears near the $\mathbf{K}$ points while gapless excitations emerge at $\mathbf{\Gamma}$ points, consistent with the previous study~\cite{sheng2025possible}. This gapless mode at $\mathbf{\Gamma}$ points is consistent with the Goldstone mode arising from the spontaneously broken U(1) symmetry associated with the uniform quadrupolar order. These low-energy excitations are also consistent with recent experimental observations~\cite{huang2025universal}. The gapped mode at $\mathbf{K}$ points and gapless mode at $\mathbf{\Gamma}$ points provide characteristic dynamical signatures of the quadrupolar spin superfluid phase. 

\section{Summary}
\label{sec:summary}
Motivated by the recent discovery of various candidate SS materials described by effective spin-1 models, we numerically investigate the spin-1 anisotropic Heisenberg model on the triangular lattice, using DMRG methods. By tuning the single-ion anisotropy $D_z$ and applied magnetic field $h_{z}$, we identify dipolar YSS and VSS phases, a quadrupolar SS, and a quadrupolar spin superfluid phase. In addition, we identify a UUD state stabilized at intermediate magnetic field, as well as a polarized phase at sufficiently high fields. By mapping out the global phase diagram as a function of $D_z$ and $h_z$, we determine the phase boundaries and characterize the different phases through their dipolar and quadrupolar order parameters and spin superfluid stiffness. Furthermore, we study the dynamical spin structure factor across the phase diagram and identify characteristic spectral signatures of different phases. We show that dipolar SS phases at small $D_z$ exhibits gapless Goldstone modes near the $\mathbf{K}$ points in both transverse component $\chi^{xy}(\mathbf{k}, \omega)$ and longitudinal component $\chi^z(\mathbf{k}, \omega)$ of the dynamical spin structure factor. In contrast, in the quadrupolar SS at large $D_z$, the gapless Goldstone excitation is only visible in $\chi^z(\mathbf{k}, \omega)$. In addition, the high-field quadrupolar spin superfluid shows gapped excitations near $\mathbf{K}$ points and gapless mode at $\mathbf{\Gamma}$ points in $\chi^z(\mathbf{k}, \omega)$. These features distinguish the uniform quadrupolar spin superfluid from the translational-symmetry-breaking quadrupolar SS. In addition to the Goldstone excitations, roton-like minima are observed near the $\mathbf{M}$ points for both the dipolar YSS at small $D_{z}$ and the quadrupolar SS at large $D_{z}$, but only under weak magnetic fields.

The phase diagram near $D_{z}= 0$ may be relevant to K$_2$Ni(SeO$_3$)$_2$~\cite{li2023k2ni,kim2026spin}, and our predictions—including the magnetic field-induced transitions among the dipolar YSS, UUD, dipolar VSS, and polarized phases—may be tested experimentally by applying an out-of-plane magnetic field. The characteristic features in the dynamical spin structure factor across the phase diagram, including the Goldstone modes and roton-like minima, are directly accessible via spin-polarized INS and offers spectroscopic signatures to identify different SS phases.

Numerical results near $D_z = 4$ are particularly relevant to Na$_2$BaNi(PO$_4$)$_2$~\cite{li2021quantum}, where a nematic quadrupolar SS has been proposed based on the spin-1 Hamiltonian~\cite{sheng2025bose,sheng2025possible,huang2025universal} and INS experiments have reported characteristic low-energy excitation spectrum~\cite{huang2025universal,sheng2025possible}. By comparing the results obtained on various lattice sizes, we characterize the quadrupolar SS at low fields through finite superfluid stiffness, longitudinal translational symmetry breaking, and transverse quadrupolar orders that are dominant at $\mathbf{\Gamma }$ points but also finite at $\mathbf{K }$ points. Direct calculations of the longitudinal dynamical spin structure factor uncover the nature of the narrow low-energy excitations observed by INS, indicating gapless Goldstone modes near $\mathbf{K}$ points for the low-field quadrupolar SS, and gapped excitations at the $\mathbf{K}$ points as well as gapless excitations at the $\mathbf{\Gamma}$ points for the high-field quadrupolar spin superfluid. Furthermore, a roton-like minimum at the $\mathbf{M}$ points is observed at the zero field for the quadrupolar SS.

Previous studies of frustrated triangular-lattice antiferromagnets have established SS phases in a variety of spin-$\frac{1}{2}$ models, where strong quantum fluctuations play an important role in stabilizing unconventional quantum phases. In higher-spin systems approaching the classical limit, quantum fluctuations are generally reduced, raising the question of whether spin supersolidity can remain stable beyond the spin-$\frac{1}{2}$ limit. Our results demonstrate robust spin supersolidity in an extended parameter regime of the spin-1 triangular antiferromagnetic Heisenberg model, establishing higher-spin triangular antiferromagnets as promising platforms for realizing both dipolar and quadrupolar spin supersolidity. It will be interesting to further explore whether related SS phases can be stabilized in triangular-lattice compounds with even larger effective spins, such as Na$_2$BaMn(PO$_4$)$_2$~\cite{zhang2024successive,Biniskos2025spin} and K$_2$Mn(SeO$_3$)$_2$~\cite{zhu2026dynamics,wang2026directional} with an effective $S=\frac{5}{2}$. Our results may also provide potential guidance in searching for experimental signals of spin supersolidity in spin-1 bilayer systems of K$_2$Ni$_2$(SeO$_3$)$_3$~\cite{li2024magnetism,yue2024magnetic} and Rb$_2$Ni$_2$(SeO$_3$)$_3$~\cite{li2025differences,chen2026nmr}. Future studies may include interlayer couplings in bilayer triangular-lattice Heisenberg models.

\begin{figure}
\centering
\includegraphics[width=1\linewidth]{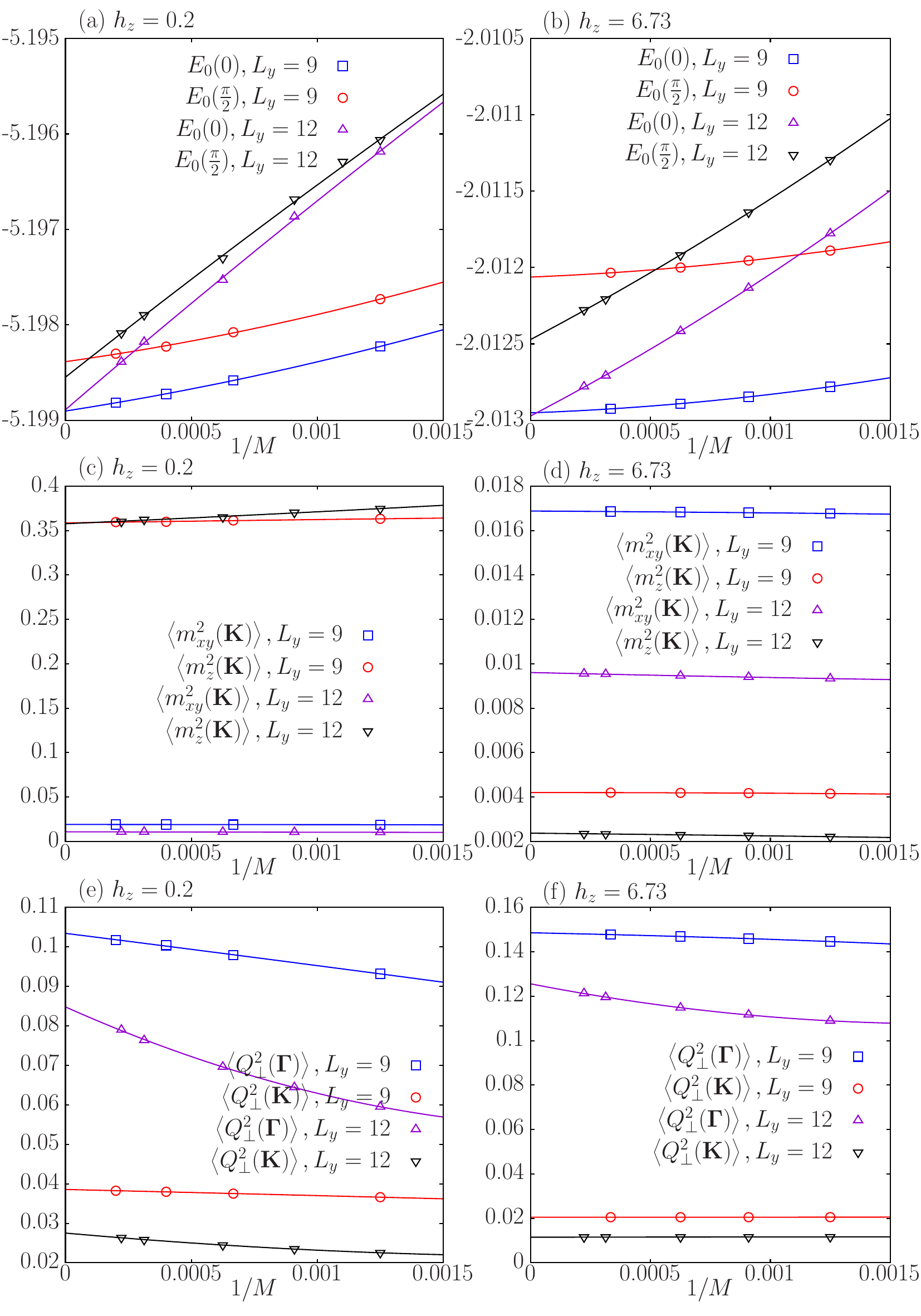}
\caption{The finite bond-dimension extrapolation of ground-state energy per site and various order parameters at $D_{z}=4$ for $L_{y}=9$ and $12$. For smaller $L_{y}$, results obtained with the largest $M$ are used instead of the extrapolated values because the convergence is much quicker with increasing $M$. The phase at $h_{z}=0.2$ refers to the quadrupolar SS and the phase at $h_{z}=6.73$ refers to the quadrupolar spin superfluid, where the extrapolated results are shown in Fig.~\ref{Fig_Dz4_order} (b) and (a), respectively.}
\label{Fig_supp_scaling_M}
\end{figure}

S.M. is financially supported by JSPS KAKENHI No. 24K00576 from MEXT, Japan. Numerical calculations were completed in part with resources provided by the HOKUSAI supercomputer at RIKEN with Project ID No. RB250023. The numerical DMRG code is implemented using the ITensor library~\cite{itensor}.

%
%

\appendix
\section{Numerical convergence and finite bond dimension extrapolation}
\label{Apendix_convergence}

\begin{figure*}
\centering
\includegraphics[width=1\linewidth]{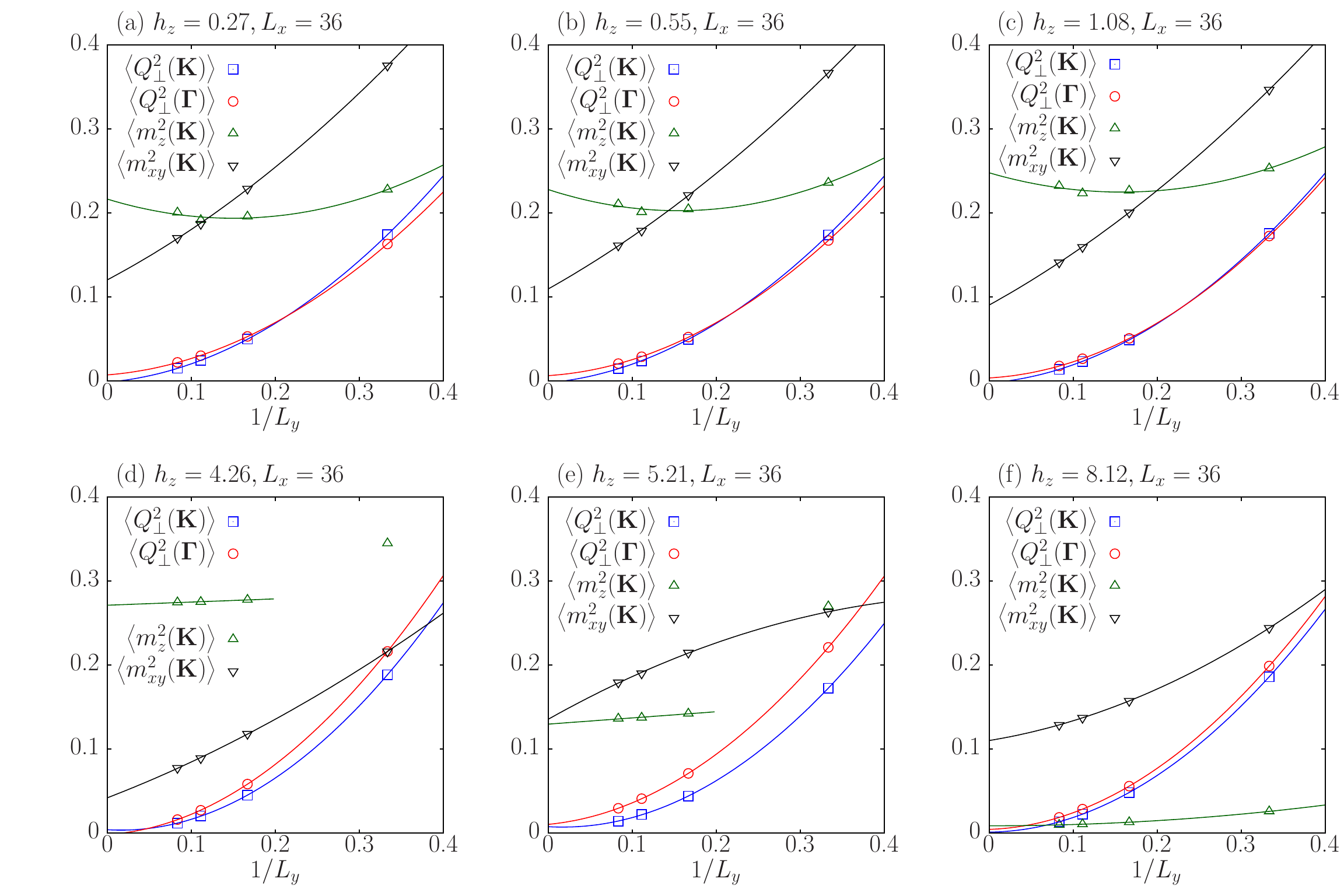}
\caption{The finite-size scaling of the order parameters obtained for various $h_{z}$ at $D_{z}=0$. The phases at $h_{z}=0.27$, $0.55$, and $1.08$ refer to the dipolar YSS. The phases at $h_{z}=4.26$, $5.21$ and $8.12$ refer to the dipolar VSS.}
\label{Fig_supp_scaling_size}
\end{figure*}

We examine the convergence of the ground-state energy per site with respect to the bond dimensions $M$. In Figs.~\ref{Fig_supp_scaling_M} (a) and (b), we show the ground-state energy per site $E_0 (0)$ as a function of the inverse bond dimensions $1/M$ for two representative points in the quadrupolar SS and quadrupolar spin superfluid phases at $D_{z}=4$, respectively. The energies converge smoothly with increasing $M$, and the extrapolated energy from a second-order polynomial fit is close to the energy obtained at the largest $M$, which indicates good convergence of the ground state.

Furthermore, we perform the same finite bond-dimension extrapolation for the order parameters and for the superfluid stiffness, the latter of which is evaluated from the energy difference $\Delta E_0(\pi/2) = E_0(\pi/2) - E_0(0)$. As shown in Figs.~\ref{Fig_supp_scaling_M} (a) and (b), $E_0 (\pi /2)$ scales similarly to $E_0 (0)$. The extrapolated energy differences for $h_{z}=0.2$ and $6.73$ are used in Fig.~\ref{Fig_Dz0} (c) and Fig.~\ref{Fig_Dz4_order} (c) of the main text, respectively. As shown in Figs.~\ref{Fig_supp_scaling_M} (c) and (d), the dipolar orders remain almost the same for different $M$. In contrast, the quadrupolar orders in both the quadrupolar SS and quadrupolar spin superfluid at $D_{z}=4$ increases with the increase of $M$, as shown in Fig.~\ref{Fig_supp_scaling_M} (e) and (f), respectively. We keep bond dimensions up to $M = 5000$, which allows reliable extrapolation of the order parameters and ground-state energies in the parameter regimes considered.

We note that on wider cylinders of $L_y \geq 9$, the numerical results become less stable near $D_z \approx 3$ at small $h_z$. This may be because the ground state in this parameter regime is close to the phase boundary of a stripe ordered state. Therefore, the ground state in this parameter regime is determined primarily from the results for $L_y = 6$ where convergence can be achieved.

\section{finite-size scaling of order parameters at $D_{z}=0$}
\label{Apendix_finite_size}

After obtaining the order parameters and superfluid stiffness extrapolated to the $M \to \infty$ limit for each system size, we perform finite-size scaling to obtain the results in the thermodynamic limit. Additional finite-size scaling results for various $h_{z}$ at $D_{z}=0$ are shown in Fig.~\ref{Fig_supp_scaling_size}. Most quantities are extrapolated by the second-order polynomial fits, except for $\left\langle m^2_z(\mathbf{K}) \right\rangle$ in Figs.~\ref{Fig_supp_scaling_size} (d) and (e), where the results obtained with $L_{y}=3$ are neglected because of their pronounced finite-size effect. Linear fits are used in these cases. The extrapolated order parameters in the thermodynamic limit are used to determine the nature of the ground states. The shaded regimes in Fig.~\ref{Fig1_phase_diagram} (a) are determined by the vanish of quadrupolar orders in the thermodynamic limit.

\begin{figure*}
\centering
\includegraphics[width=0.8\linewidth]{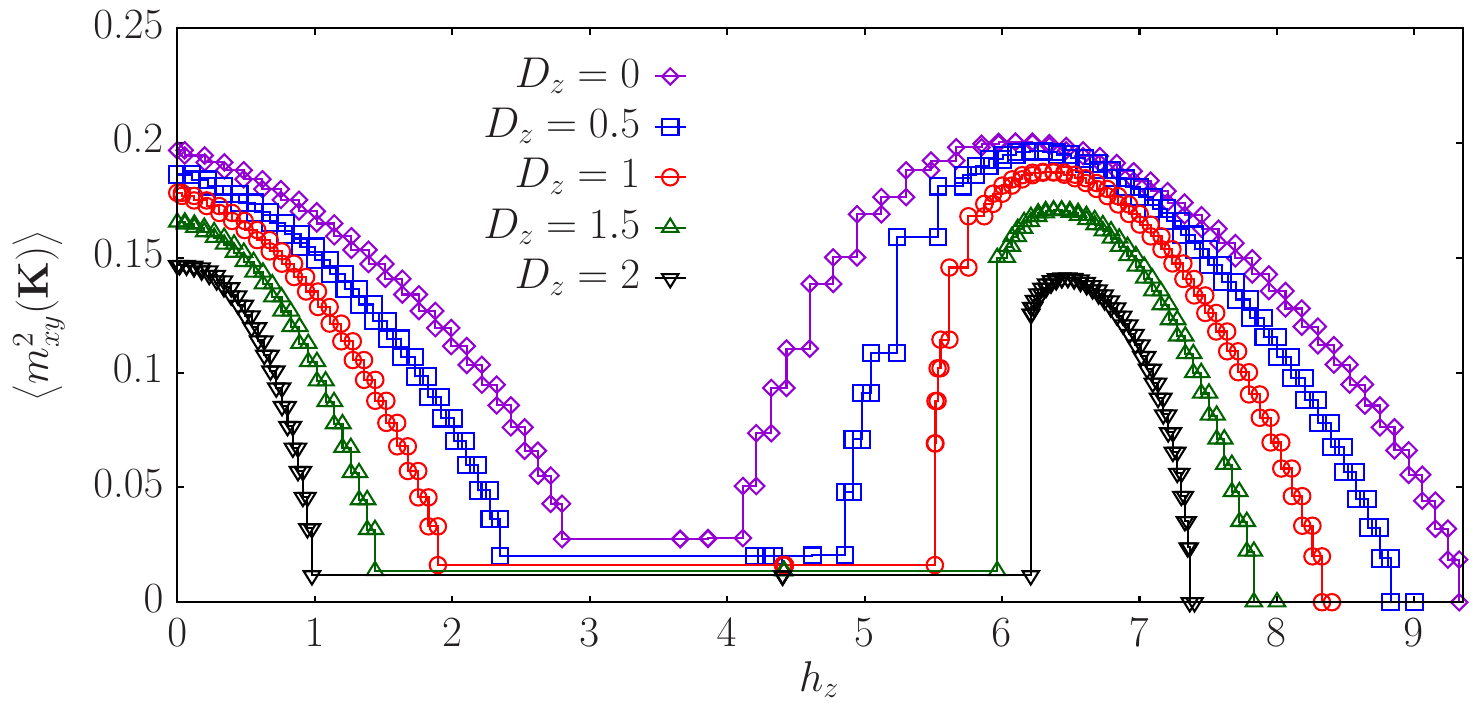}
\caption{$\left\langle m_{xy}^{2} (\mathbf{K}) \right\rangle$ as a function of $h_{z}$ for various $D_{z}$. The results are obtained on the lattice of $N = 24 \times 6$.}
\label{Fig_supp_fix_lattice}
\end{figure*}

\section{Nature of transitions from the UUD phase to the Dipolar VSS phase}
\label{Apendix_fixed_lattice}

As shown in Fig.~\ref{Fig_supp_fix_lattice}, at small $D_{z}<0.5$ the transverse dipolar order $\left\langle m_{xy}^{2} (\mathbf{K}) \right\rangle$ increases smoothly from the UUD phase to the dipolar VSS phase, which is consistent with a continuous phase transition. However, for larger $D_{z}>1.5$ a sudden jump in $\left\langle m_{xy}^{2} (\mathbf{K}) \right\rangle$ is observed from the UUD phase to the dipolar VSS phase, indicating a first-order phase transition. Sudden jumps of $\left\langle Q_{\perp }^{2} ( \mathbf{\Gamma}) \right\rangle$ from the UUD phase to the quadrupolar spin superfluid for $D_{z}>2.5$ are also observed to indicate a first-order phase transition, which is consistent with previous studies in the large $D_{z}\approx 4$~\cite{sheng2025possible,huang2025universal}.

\section{Real-space distribution of $\left \langle S^{z}_{i} \right \rangle$ for various phases}
\label{Apendix_real_space}
\begin{figure*}
\centering
\includegraphics[width=0.95\linewidth]{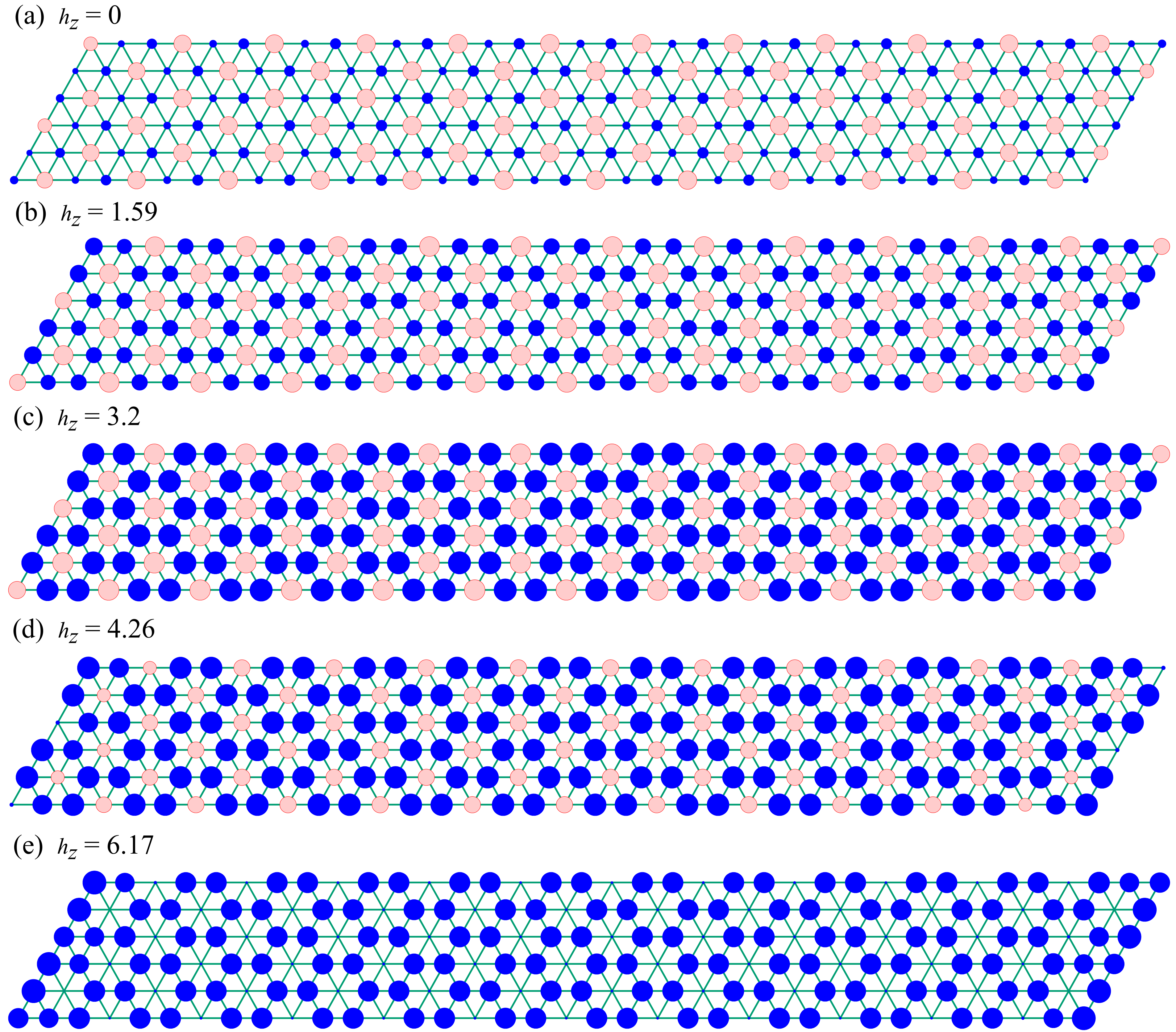}
\caption{The real-space distribution of $\left \langle S^{z}_{i} \right \rangle$ for various $h_{z}$ at $D_{z}=0$. The blue solid circles refer to positive $\left \langle S^{z}_{i} \right \rangle$, and red shaded circles refer to negative $\left \langle S^{z}_{i} \right \rangle$. The radius is proportional to the magnitude of $\left \langle S^{z}_{i} \right \rangle$. The blue solid circles in the center of panel (c) have $\left\langle S_{i}^{z}\right\rangle \approx 0.92$. All results are obtained on the lattice of $N = 36 \times 6$.}
\label{Fig_supp_real_space_Dz0}
\end{figure*}

We show additional results of the real-space distribution of $\left \langle S^{z}_{i} \right \rangle$ in Fig.~\ref{Fig_supp_real_space_Dz0} for various $h_{z}$ at $D_{z}=0$. The states at (a) $h_{z}=0$, (b) $1.59$, (c) $3.2$, (d) $4.26$, (e) $6.17$ correspond to the dipolar YSS, dipolar YSS, UUD, dipolar VSS, and dipolar VSS phases, respectively. In all cases, the real-space spin configurations exhibit three-sublattice modulations, corresponding to peaks in $\left\langle m_{z}^{2} (\mathbf{k}) \right\rangle$ at the $\mathbf{K}$ points.

\bibliography{Spin1_supersolid}

\end{document}